\documentclass{article}

\usepackage{arxiv}

\usepackage[utf8]{inputenc} 
\usepackage[T1]{fontenc}    
\usepackage{hyperref}       
\usepackage{url}            
\usepackage{booktabs}       
\usepackage{amsfonts}       
\usepackage{nicefrac}       
\usepackage{microtype}      
\usepackage{lipsum}
\usepackage{color}
\usepackage{algorithm}
\usepackage{algorithmic}
\usepackage{amsmath,amsfonts,amssymb}
\usepackage{subfigure}
\usepackage{lineno}
\usepackage{graphicx}
\usepackage{bm}
\graphicspath{{Figures/}}
\usepackage{subfigure}

\title{Sparse sensor reconstruction of vortex-impinged airfoil wake with machine learning}

\author{
Yonghong Zhong$^{[1,*]}$, Kai Fukami$^{[1]}$, Byungjin An$^{[2]}$, Kunihiko Taira$^{[1]}$\\
1. Department of Mechanical and Aerospace Engineering, \\
University of California, Los Angeles, CA 90095, USA\\
2. Fundamental Technologies, R\&D Department, Ebara Corporation, Tokyo 144-8510, Japan\\
Corresponding author: yhzhong@g.ucla.edu
}

\begin{document}
\maketitle

\begin{abstract}
Reconstruction of unsteady vortical flow fields from limited sensor measurements is challenging. 
We develop machine learning methods to reconstruct flow features from sparse sensor measurements during transient vortex-airfoil wake interaction using only a limited amount of training data.
The present machine learning models accurately reconstruct the aerodynamic force coefficients, pressure distributions over airfoil surface, and two-dimensional vorticity field for a variety of untrained cases. 
Multi-layer perceptron is used for estimating aerodynamic forces and pressure profiles over the surface, establishing a nonlinear model between the pressure sensor measurements and the output variables.
A combination of multi-layer perceptron with convolutional neural network is utilized to reconstruct the vortical wake.
Furthermore, the use of transfer learning and long short-term memory algorithm combined in the training models greatly improves the reconstruction of transient wakes by embedding the dynamics. The present machine-learning methods are able to estimate the transient flow features while exhibiting robustness against noisy sensor measurements.
Finally, appropriate sensor locations over different time periods are assessed for accurately estimating the wakes.
The present study offers insights into the dynamics of vortex-airfoil interaction and the development of data-driven flow estimation.

\keywords{Vortex-airfoil interaction, Machine learning, Flow reconstruction}
\end{abstract}

\section{Introduction}
{
Vortex-airfoil interaction is ubiquitous around fluid-based systems, including aircraft~\cite{jones2022physics,Jeff2021,pfnur2019leading,iannelli2019worst}, wind turbines~\cite{scherl2020geometric}, and pumps~\cite{liu2018core,LiuJFE}. 
Such interactions can cause unsteady loading, fatigue, and structural damage to these systems.
For analyzing vortex-airfoil interactions, it is useful to assess the state of the flow from sparse measurements for understanding the governing dynamics~\cite{halder2020deep}, prediction of flow disturbance~\cite{Girguis2022}, and performing the wake flow control~\cite{herrmann2022gust}.
However, it is challenging to identify vortical structures during the vortex-airfoil interactions from sparse measurements due to its strong nonlinear dynamics and the high-degree of freedom required to describe the vortical flows.

A number of studies have examined sparse state estimation for aerodynamics.
In particular, linear techniques have been studied over the last several decades.
For instance, gappy proper orthogonal decomposition~\cite{ES1995} has been considered to obtain dominant flow features from spatially incomplete and sparse data sets~\cite{BDW2004}.
Focusing on the characterization of flows and boundary layers near body surface, the applications of four-dimensional variational method~\cite{bewley2001dns}, linear stochastic estimation~\cite{adrian1988stochastic}, and Kalman filters~\cite{colburn2011state} have also been explored.
However, these techniques are constrained by their linear formulations, which poses challenges when the applications involve strongly nonlinear dynamics.

To overcome such limitations, nonlinear machine learning approaches have been considered as a promising approach in analyzing fluid flows from sparse information.
Nonlinear machine learning techniques have shown to be useful in estimating and modeling high-dimensional flow~\cite{Steven2020}.
For example, Pawar et al.~\cite{pawar2021physics,pawar2022multi} applied a physics-guided machine-learning framework to estimate the lift coefficient of a variety of airfoils.
Hui et al.~\cite{hui2020fast} utilized a signed distance function-assisted convolutional neural network (CNN) to predict the pressure distribution over an airfoil surface.
For flow field reconstructions, Erichson et al.~\cite{erichson2020} proposed a shallow decoder based on multi-layer perceptron (MLP) for a circular cylinder wake, the sea surface temperature, and forced isotropic turbulence. 
Fukami et al.~\cite{fukami2021global} proposed a CNN-based method to reconstruct the global turbulent flow field from sparse sensors that can be in motion or change in numbers.
In addition to the aforementioned efforts, there are various machine-learning-based flow reconstruction techniques based on super-resolution analysis~\cite{FFT2019,GANSRkim2021,CZXG2019}.

However, there are issues with utilizing nonlinear machine learning techniques for estimating unsteady fluid flows from limited sensor measurements.
{The most outstanding issue is the computational costs for using machine learning models are expensive. 
For neural network-based models with low-dimensional inputs to high-dimensional outputs, an enormous number of interior parameters (weights) are required. 
To determine the internal parameters, generally, thousands of flow (or sensor) snapshots are required, which causes a large computational burden in terms of both training costs and data storage.
In our case, if a variety of unsteady flow fields is needed to be accurately reconstructed, storage and computing costs can rise significantly if the problem is approached naively.}
From this aspect, it is crucial to develop a method that can qualitatively reconstruct a flow field with a small amount of training data and a reduced number of tuning parameters.
{In addition, generalizable models promote a reduction in cost.
Most machine learning models can only be used for specific flow fields, for example, a single model trained with a laminar flow may not be applicable to use to reconstruct turbulent flow fields. In fact, the data used for testing needs to be similar to the training data to achieve accurate results. If we need to consider different flows over a vast parameter space, it is almost impossible to perform experiments or simulations for each and every case. In this regard, the diversity of the training data needs to be considered so that a single model can effectively predict unsteady flow fields over a large range of parameters.}

In this study, we aim to develop machine learning methods that reconstruct dominant wake features from limited sensor measurements and a small set of training data sampled over a vast parameter space. 
Because the disturbance vortex can be of any size, strength, or position from the airfoil, a very large parameter space is needed to be explored to capture the complex vortex-impinged airfoil wake dynamics. 
In this case, the amount of data can be tremendously large. Instead of naively training machine learning models with all parameter combinations, we develop models that are trained with a few cases in the parameter space and use the models to estimate unseen cases.
For the machine learning methods, we choose a multi-layer perceptron (MLP) to model the nonlinear relationship between the low-dimensional sensors inputs and the outputs, including the lift coefficient, drag coefficient, and surface pressure coefficient.
Moreover, combining the convolutional neural networks and MLP allows the reconstruction of the vorticity field over time with modest computational costs. 
The transfer learning and long-short term memory further help in incorporating the dynamics of the transient flow, which reduces the required training data and improves the flow estimation.
The current model is robust for a variety of wake scenarios separate from the training data. We also assess the influence of sensor numbers and placement on flow estimation.

The present paper is organized as follows. The problem setup and data compilation are discussed in section~\ref{sec2}. 
Flow physics of vortex-airfoil wake interactions are presented in section~\ref{sec:physics}.
Machine learning techniques utilized in this study are introduced in section~\ref{sec:ML}.
Results and discussion of machine learning-based flow reconstruction are presented in section~\ref{sec:res}.
Concluding remarks are provided in section~\ref{sec:conc}.}

\section{Data compilation}
\label{sec2}

\begin{figure}
  \centering
    \includegraphics[width=0.7\textwidth]{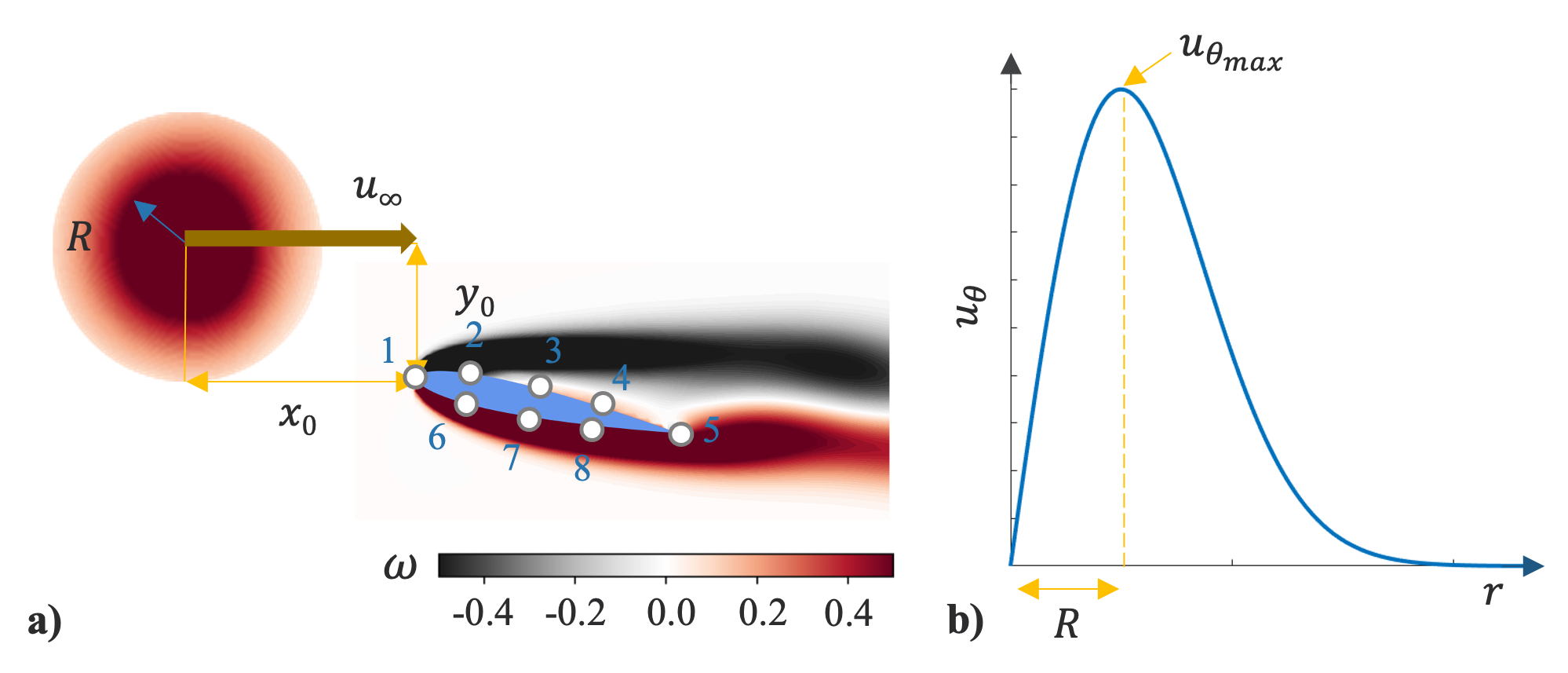}
    \caption{a)~The size and position of the vortical disturbance, and 8 uniform sensors are distributed on the airfoil surface; b)~The velocity profile of the disturbance vortex.
    }
    \label{fig_c1}
\end{figure}

\begin{figure}[t]
  \centering
    \includegraphics[width=0.9\textwidth]{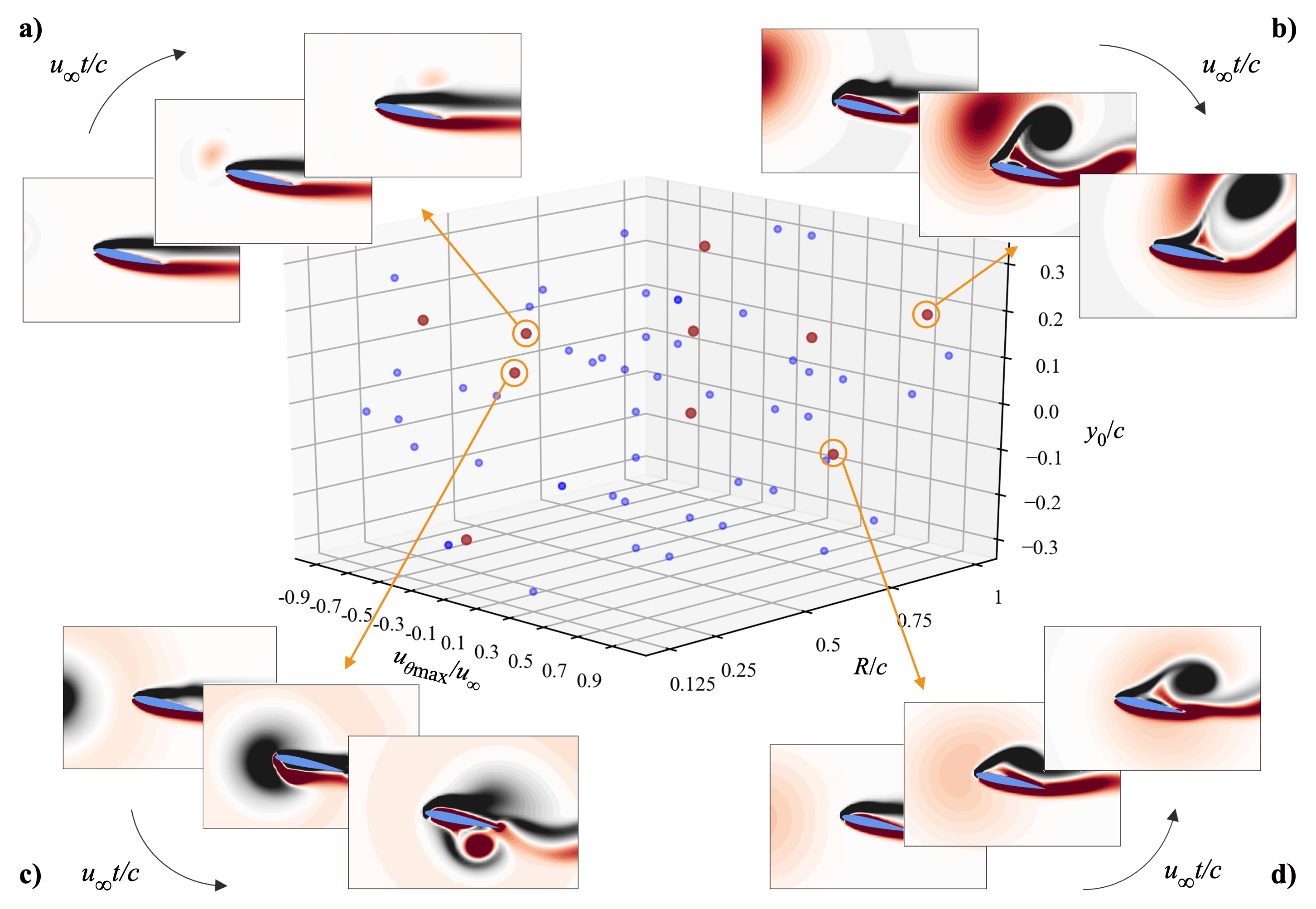}
    \caption{Randomly distributed 50 training cases (blue) and 10 test cases (red). Example test cases shown: a)~$(u_{\theta \rm{max}}/u_{\infty}, R/c, y_{0}/c)=(0.14, 0.15, 0.15)$, b)~$(u_{\theta \rm{max}}/u_{\infty}, R/c, y_{0}/c)=(0.78, 0.99, 0.18)$, c)~$(u_{\theta \rm{max}}/u_{\infty}, R/c, y_{0}/c)=(-0.80, 0.61, 0.08)$, d)~$(u_{\theta \rm{max}}/u_{\infty}, R/c, y_{0}/c)=(0.35, 0.95, -0.15)$.
    }
    \label{fig_distribution}
\end{figure}

The present objective is to develop a robust machine-learning model for highly disturbed flows around an airfoil from sparse pressure sensors and limited training data.
Here, we consider transient flow over a NACA 0012 airfoil at an angle of attack of $\alpha=12^{\circ}$ experiencing various types of vortical disturbances at a chord-based Reynolds number $Re \equiv u_{\infty} c/{\nu}_{\infty}=400$ and a Mach number $M_{\infty}\equiv u_{\infty}/a_{\infty}=0.1$. Here, $u_{\infty}$ is the free-stream velocity, $c$ is the chord length, ${\nu}_{\infty}$ is the kinematic viscosity, and $a_{\infty}$ is the freestream sonic speed. 
The simulated flows have been verified and validated with previous studies~\cite{kurtulus2015unsteady,liu2012numerical,di2018fluid}.

The compressible flow solver CharLES~\cite{Bres} is used to simulate the transient flows over the airfoil. For the present vortex-airfoil interaction problem, a single vortical disturbance is initially introduced upstream of the airfoil. 
This disturbance vortex is given as a compressible Taylor vortex~\cite{taylor1918dissipation}, described by
\begin{equation}
    u_{\theta}=u_{\theta \rm{max}}{\dfrac{r}{R}}{\rm exp}\left[\dfrac{1}{2}\left({1-\dfrac{r^2}{R^2}}\right)\right] ,
\end{equation}
where $R$ is the radius, and $u_{\theta \rm{max}}$ is the maximum rotational velocity of the vortex, as shown in figure~\ref{fig_c1}.
The vortex is initially introduced at $(x_0,y_0)$ with $x_0=-2c$.

The present vortex-airfoil interaction problem exhibits a variety of flow patterns, as shown in figure~\ref{fig_distribution}.
A strong disturbance vortex produces strong unsteadiness in the flow field, and the larger the vortex is, the larger the region it influences. 
Apart from the radius and the strength, a vortex can either hit the airfoil at the leading edge and thus incite large fluctuations or pass through the airfoil without causing dramatic changes to the flow or aerodynamic characteristics. Detailed discussion on the flow is offered in section~\ref{sec:physics}.
The present study examines whether the flow field generated over the wide parameter space can be recovered with the machine-learning model trained with only a very few cases.

In the present study, we choose eight sensors distributed on both sides of the airfoil surface to capture the vortex passing around an airfoil, as shown in figure~\ref{fig_c1}. 
These sensors are labeled $1$ to $8$, with the respective $x$-locations of the sensors being $(0.00, 0.26, 0.48, 0.72, 0.99, 0.23, 0.46, 0.71)c$.
Three parameters that describe the disturbance vortex are maximum rotational velocity ($u_{\theta \rm{max}}$), the radius ($R$), and the initial vertical location ($y_0$). The training data sets are comprised of $u_{\theta \rm{max}}/u_{\infty} \in [-0.9, -0.7, -0.5, -0.3, -0.1, 0.1, 0.3, 0.5, 0.7,\\
0.9]$, $R/c\in [0.125, 0.25, 0.5, 0.75, 1]$, and $y_{0}/c\in [-0.3, -0.1, 0, 0.1, 0.3]$, respectively. 
Here, the positive value of $u_{\theta \rm{max}}/u_{\infty}$ indicates a counterclockwise rotation.
The maximum rotational velocity of the vortex $u_{\theta \rm{max}}$ covers a range from $0.1 u_{\infty}$ to $0.9 u_{\infty}$.
The choices for the vortex radius $R$ and $y_0$ are carefully determined so that vortices can pass over or below the airfoil while significantly influencing the airfoil wake. 
In section~\ref{sec:res}, we consider 25, 50, and 100 training cases out of the vast combinations of parameters, then test the models with untrained cases. 
Parameter combinations of test cases are randomly chosen over the aforementioned ranges. 
Note that the training data is a small proportion compared to the whole combinations of parameters.
There are no test cases overlapping with the training cases.

For each case, we collect 500 snapshots of the flow field for $u_{\infty}t/c \in [0.85,5.1]$, which reflects the process from the vortex approaching the airfoil to moving away from the tailing edge. Here, $u_{\infty}t/c=0$ refers to the initial time at which the vortex is at $x_{0}/c=-2$.
The snapshots at $u_{\infty}t/c=[0,0.85]$ are not used in the present analysis to remove the start-up period of the simulation.
For a single parameter set $(u_{\theta \rm{max}}/u_{\infty},R/c,y_{0}/c)$, the data sizes of aerodynamic force coefficients, pressure over surface, and two-dimensional vorticity field data amount to approximately 1MB, 15MB, and 500MB, respectively. 
If we use 100 training cases with all 500 snapshots of two-dimensional wake data, the training data size becomes approximately 50GB for a single machine learning model, which is quite large with respect to storage and computation.

\section{Flow physics}
\label{sec:physics}

{The present vortex-impinged airfoil wake exhibits rich dynamics influenced by the vortex velocity, size, and position.
In this section, we present the flow physics induced by a variety of vortex disturbances.

\begin{figure}[t]
  \centering
    \includegraphics[width=1\textwidth]{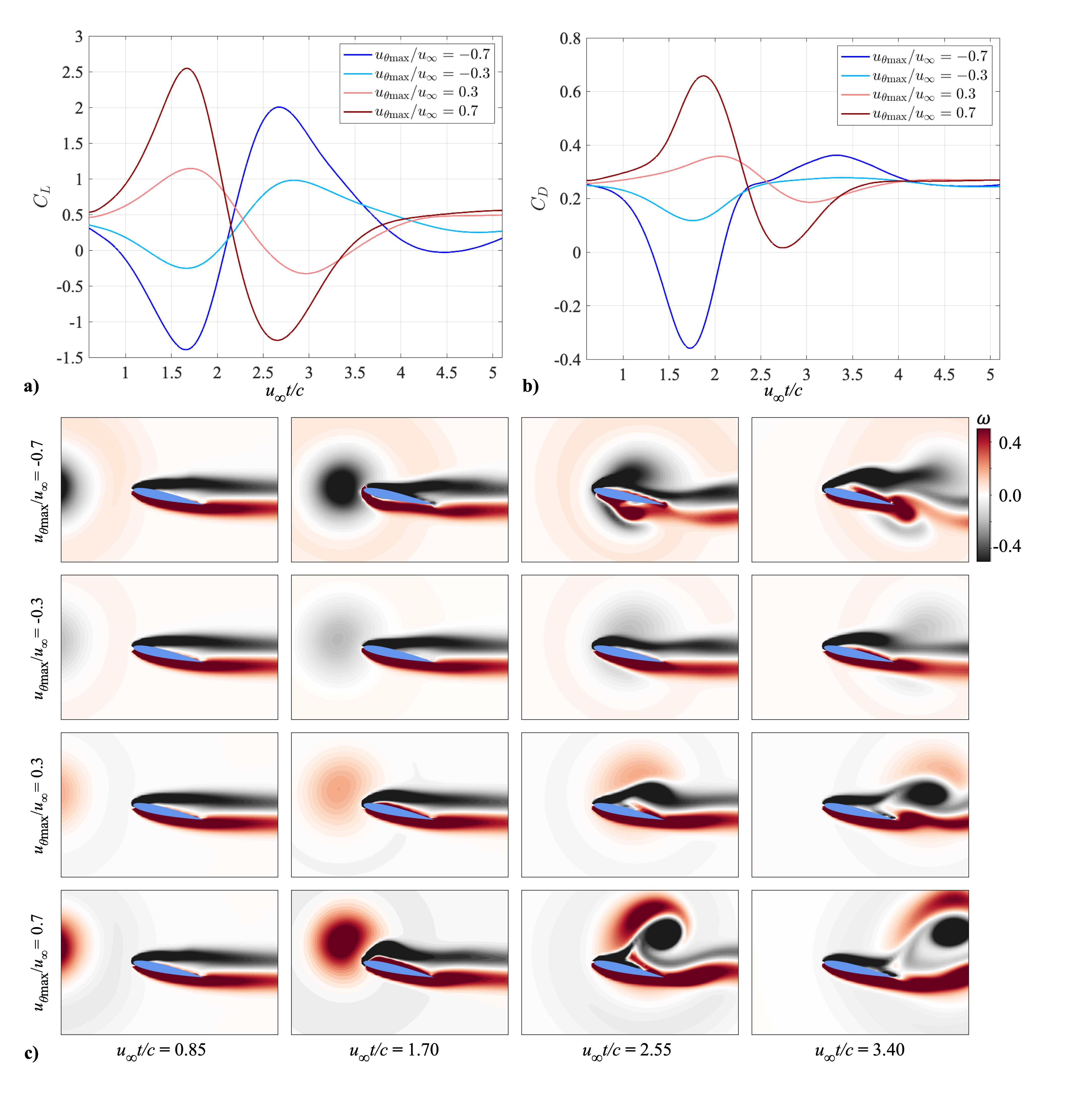}
    \caption{Effect of the largest rotational velocity of vortical disturbance. a)~lift coefficients, b)~drag coefficients, and c)~vorticity fields for vortical disturbances of $(R/c, y_{0}/c)=(0.5, 0.1)$ and $u_{\theta \rm{max}}/u_{\infty}=-0.7,-0.3,0.3$, and $0.7$.
    }
    \label{strength}
\end{figure}

The maximum rotational velocity of the vortex disturbance is one of the most important characteristics affecting the vortex-airfoil interaction. Here, we investigate the influence of vortex largest velocity on $C_L$, $C_D$, and vorticity fields when $(R/c,y_{0}/c)=(0.5,0.1)$.
As depicted in figure~\ref{strength}a) and b), a positive (counterclockwise) vortex generally induces a transient increase in $C_L$ and $C_D$ when it impinges on the leading edge of the airfoil. 
A secondary negative peak is then introduced when the center of the vortical disturbance passes the center of the airfoil. 
A similar but reversed trend is observed for a negative (clockwise) vortex. 
The initial decrease in lift is followed by the vortex tail-induced lift increase.

For a positive vortex with two different magnitudes of the vortex rotational velocity, the first peaks of~$C_L$ are reached at nearly the same time, as presented in figure~\ref{strength}a).
However, the magnitude difference causes the temporal shift for the secondary peak --- the peak with $u_{\theta_{\rm max}}/u_{\infty}=0.7$ is reached at $u_{\infty}t/c\approx 2.6$ while that with $u_{\theta_{\rm max}}/u_{\infty}=0.3$ is achieved at $u_{\infty}t/c\approx 3.0$.
This is because a stronger positive vortex produces a stronger interaction with the pre-existing negative vorticity on the suction side of the airfoil, forming a large negative vortex that detaches from the airfoil afterward. 
Similar to the positive disturbance cases, larger fluctuation induced by a stronger negative vortex gives rise to an earlier secondary peak. 
For $C_D$, we observe a similar trend of the time history to the $C_L$ for the positive disturbance, while the magnitudes of variation are much smaller than $C_L$.

\begin{figure}[t]
  \centering
    \includegraphics[width=1\textwidth]{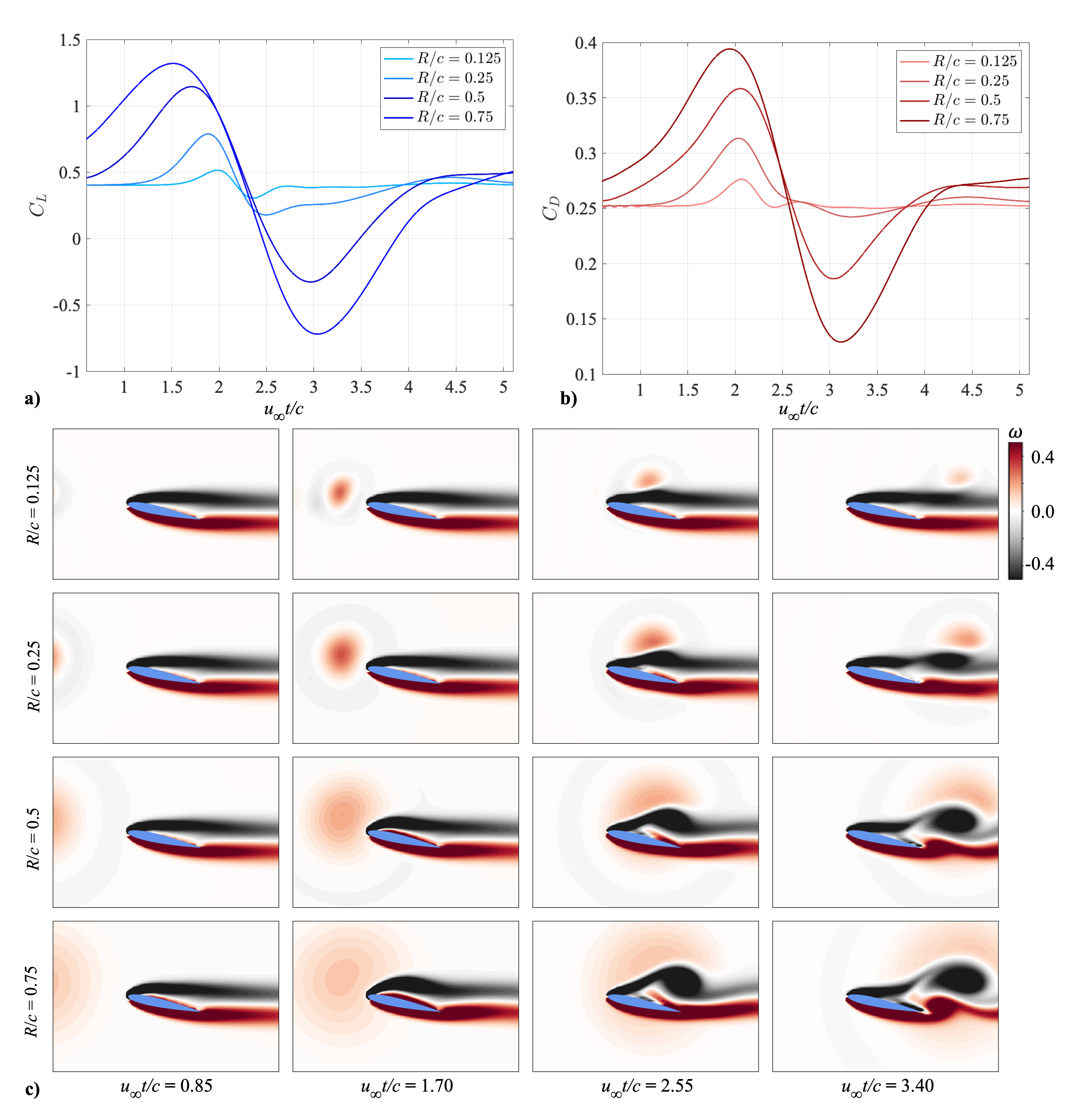}
    \caption{Effect of vortex size. a)~lift coefficients, b)~drag coefficient, and c)~vorticity fields for vortical disturbances of $(u_{\theta \rm{max}}/u_{\infty}, y_{0}/c)=(0.3, 0.1)$ and $R/c=0.125,0.25,0.5$, and $0.75$.
    }
    \label{radius}
\end{figure}

The dependence of the flow field response on the vortex size is also examined, as shown in figure~\ref{radius}.
We choose the same vortex strength and vertical position as $(u_{\theta \rm{max}}/u_{\infty}, y_{0}/c)=(0.3, 0.1)$ for comparison.
The $C_L$ and $C_D$ histories experience the same trend of first increasing and then decreasing among different vortex sizes. 
By increasing the vortex size, the first peaks of $C_L$ and $C_D$ appear earlier because a vortex with a larger radius encounters the airfoil earlier.

The changes in the vorticity fields caused by the different sizes of vortices are also presented in figure~\ref{radius}c). 
When a small-size vortical disturbance ($R/c=0.125$) impinges on the airfoil, the whole vortex passes over the suction side of the airfoil and induces mild fluctuation in the flow field. 
As the size of the vortex becomes larger, the vortex splits into two structures which advects over the suction side and the pressure side. 
The positive vorticity around the trailing edge is rolled up and interacts with the wakes, thus affecting the evolution of the wake region.

\begin{figure}
  \centering
    \includegraphics[width=1\textwidth]{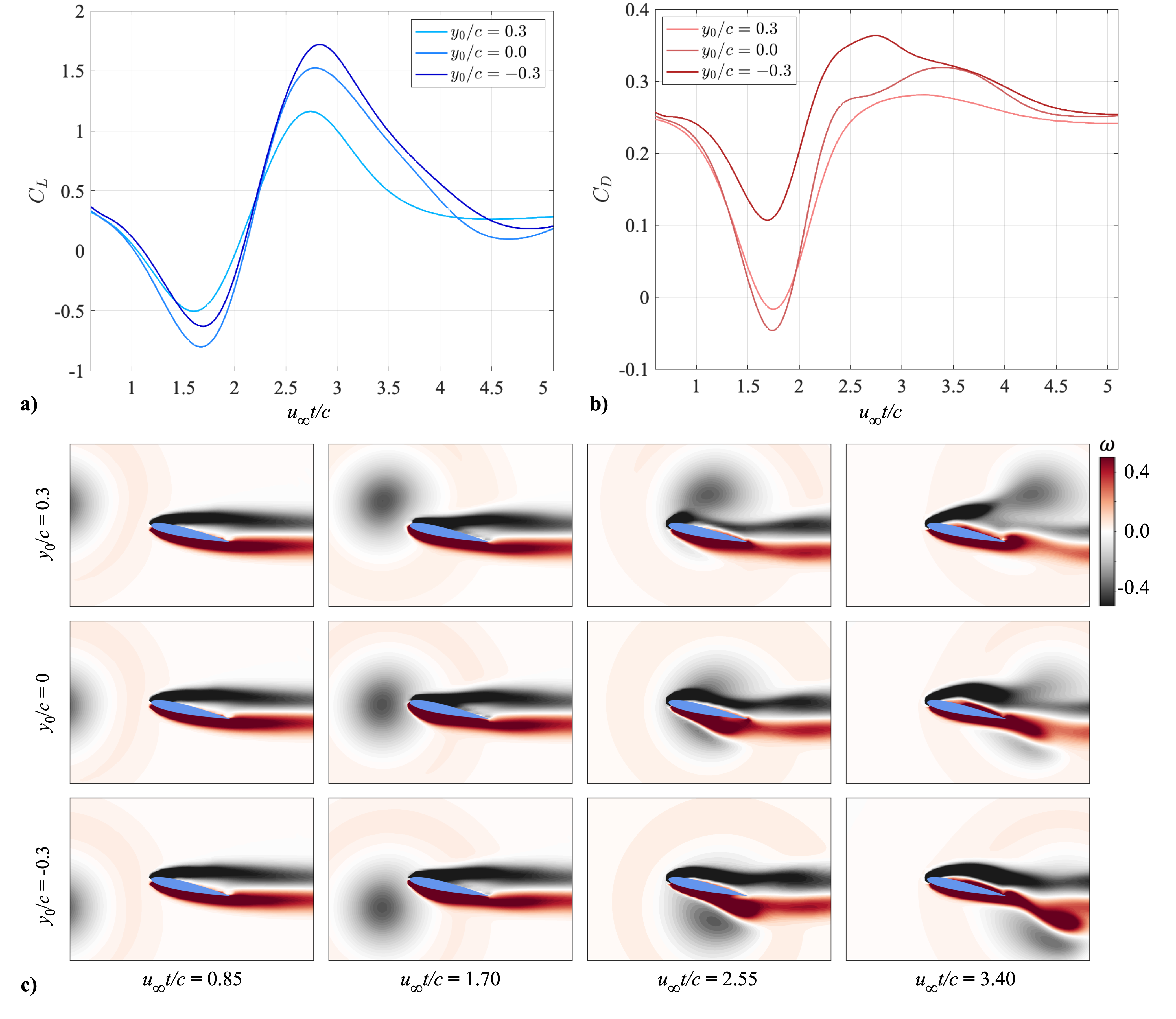}
    \caption{Effect of vortex position. a)~lift coefficients, b)~drag coefficients, and c)~vorticity fields for vortical disturbances of $(u_{\theta \rm{max}}/u_{\infty}, R/c)=(-0.5, 0.5)$ and $y_{0}/c=0.3,0$, and $-0.3$.
    }
    \label{y0}
\end{figure}

In addition to the largest velocity and the size of the vortical disturbance, the transient dynamics are also strongly influenced by whether the disturbance vortex passes above or below the airfoil.
Here, let we investigate three vertical positions of $y_{0}/c={-0.3,0,0.3}$ with a negative vortical disturbance $(u_{\theta \rm{max}}/u_{\infty}, R/c)=(-0.5, 0.5)$, as shown in figure~\ref{y0}.
For $y_{0}/c=0.3$, the disturbance passes over the airfoil, where a large portion of negative disturbance passes through the suction side of the airfoil, introducing a large jump in $C_D$ as the first peak.
For $y_{0}/c=0$, the negative vortical disturbance is split into two parts as it passes around the airfoil. 
At $u_{\infty}t/c=2.55$, the large positive vorticity attached on the pressure side of the airfoil produces the second peak in $C_L$.
For the case of $y_{0}/c=-0.3$ where the disturbance passes below the airfoil, the variation is mostly dominated by the interaction along the pressure side of the airfoil, and the drop and the increment of $C_L$ and $C_D$ occur at the same time.
}

\section{Methods}
\label{sec:ML}

{
We develop machine-learning models to estimate aerodynamic characteristics that cover a variety of force and wake dynamics from sparse sensors.
Constructing a robust model suitable for the vast parameter space in figure~\ref{fig_distribution} is challenging.
To estimate different types of nonlinear wake responses from limited training data, we consider several strategies with regard to machine-learning model design and training methods for reproducing the transient dynamics. 
For all machine-learning models used in the current study, three-fold cross-validations are performed, ensuring the convergence of the estimations in terms of data distribution.

\begin{figure}[t]
  \centering
    \includegraphics[width=0.8\textwidth]{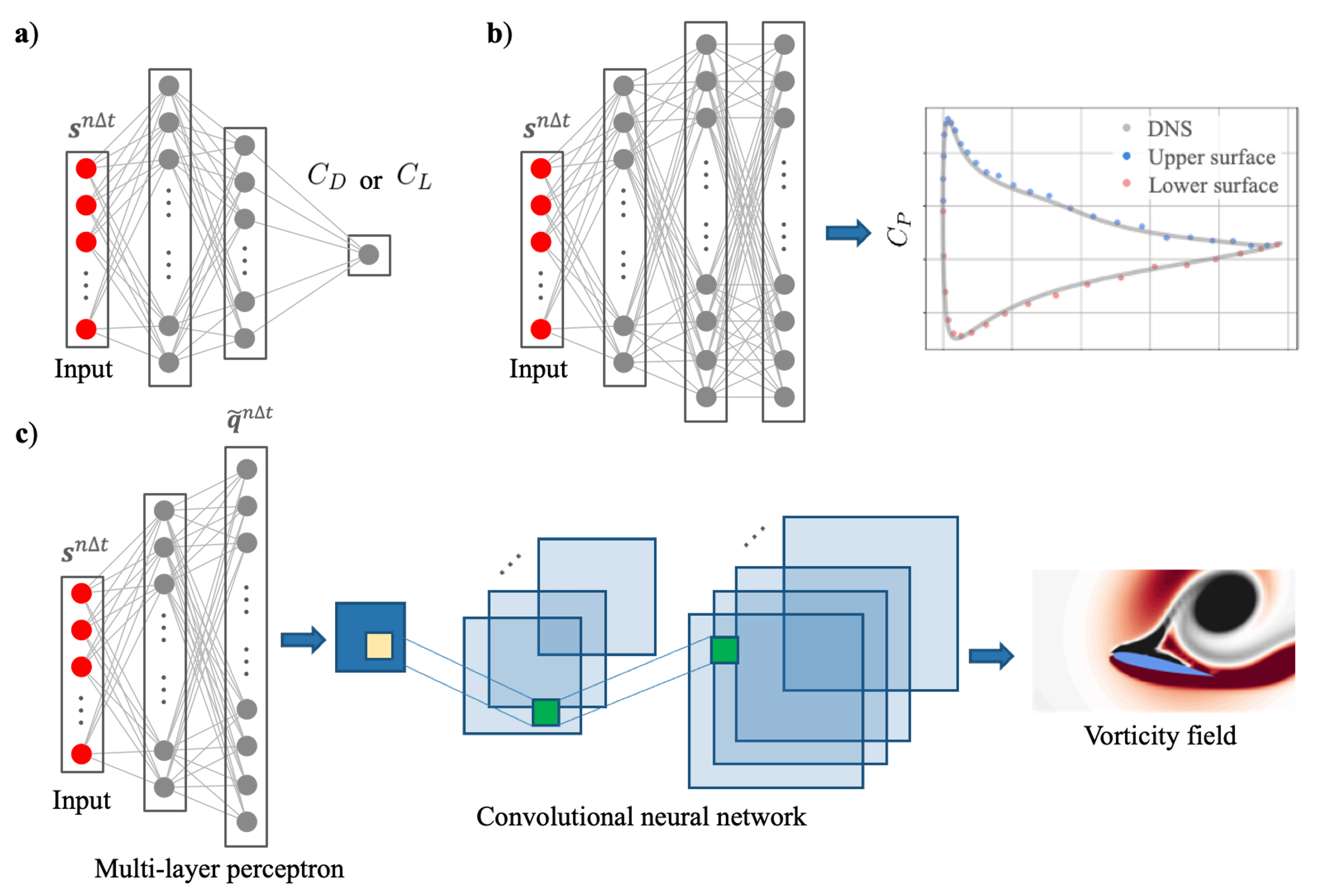}
    \caption{{
    Overview of the present estimation problems. 
    The inputs are pressure sensor measurements on the airfoil surface, outputs are
    a)~$C_D$ or $C_L$, b)~$C_P$, and c)~vorticity field.
    $C_L$, $C_D$, and $C_P$ are estimated using separate multi-layer perceptron models, vorticity field is estimated using the combination of multi-layer perception and convolutional neural network.}
    }
    \label{fig_c2}
\end{figure}

An overview of the present machine-learning-based estimation approaches is shown in figures~\ref{fig_c2}. The input is the sensor measurements ${\bm{s}}^{n\rm{\Delta} t}$ spanning over $n\rm{\Delta} t$.
We first consider a multi-layer perceptron (MLP) to build the relationship between the sensor measurements and the aerodynamic force coefficients over time. 
Since the degrees of freedom of the input and output are ${\cal O}(1)$, we can easily employ a fully-connected neural network to construct such a relationship.
Similarly, we also use a multi-layer perceptron (MLP) to estimate the pressure distribution over the airfoil surface. 
However, MLP can be challenging to use for problems with high degrees of freedom due to its fully-connected structure~\cite{fukami2021global,wu2020comprehensive}. 
To access the two-dimensional vorticity flow field (the degree of freedom $\approx{\cal O}(10^{3}-10^{4})$), a model which can effectively extract spatial information with a manageable computational cost is required.
To address this point, we incorporate a two-dimensional convolutional neural network (CNN) to provide qualitative estimations while maintaining a low computational cost.
When the MLP is coupled with the CNN, the machine-learning model can reconstruct the flow field from a limited number of sensor measurements.
Moreover, due to the transient nature of the current vortex-airfoil interaction problem, accounting for the dynamics into model construction aids in accurate estimation. 
For this reason, the long short-term memory (LSTM) algorithm~\cite{HS1997} assisted with transfer learning serves as an effective method to estimate the flow fields from time traces. 
Hence, we embed LSTM into the aforementioned MLP and MLP-CNN models. 
In what follows, we introduce the algorithms of these machine learning methods.
}

\begin{figure}
  \centering
    \includegraphics[width=0.9\textwidth]{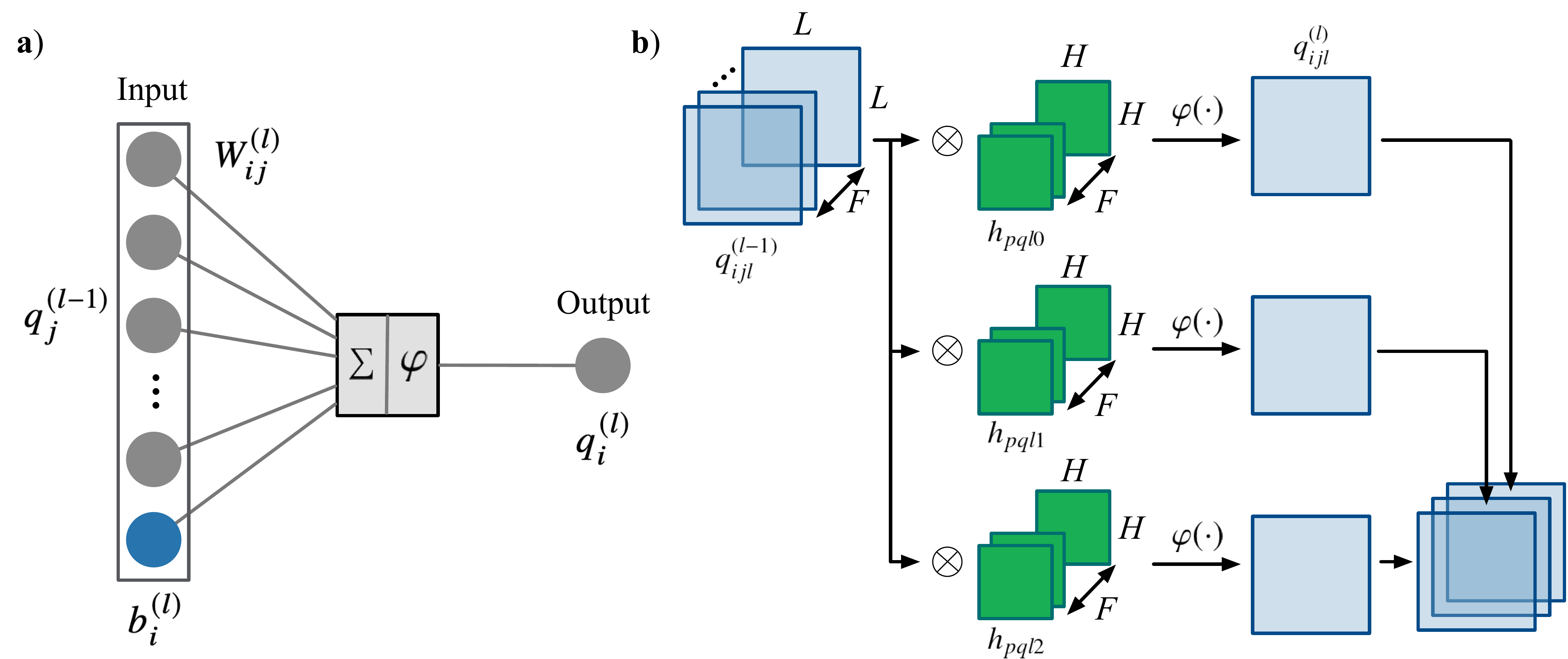}
    \caption{{
    a)~A minimum unit of perceptron. b)~Two-dimensional convolutional operation.}
    }
    \label{fig_models}
\end{figure}

\subsection{Multi-layer perceptron}
\label{sec:mlp}

In the present study, the input sensor measurements are first fed into a multi-layer perceptron (MLP)~\cite{Rumelhart}.
For the estimation of aerodynamic forces (section~\ref{sec:cdcl}) and pressure distribution over the airfoil surface (section~\ref{sec:pres}), the MLP ${\cal M}$ is used as a function approximator between the input sensor measurements ${\bm s}$ and the output variables ${\bm q}$ such that ${\bm q}\approx {\cal M}({\bm s})$. 
For the estimation of the two-dimensional vorticity field~$\omega$ (section~\ref{sec:2D}), MLP plays a role of a nonlinear function mapping the low-dimensional sensor information~${\bm s}\in \mathbb{R}^{n_{\bm s}}$ to the high-dimensional variable in the model.
In addition, we incorporate LSTM~\cite{HS1997} into the machine-learning models to capitalize on the dynamical information of sensors.

In MLP, the input at layer $(l-1)$ is multiplied by weights $\bm W$, then linearly combined, and passed through a nonlinear activation function $\varphi$ as an output to the next layer $(l)$,
\begin{equation}
    {q}^{(l)}_{i}=\varphi(\sum_{j}{W}_{ij}^{(l)}{q}^{(l-1)}_{j} + b_i^{(l)}),    
\end{equation}
where $b$ is a bias added at each layer as illustrated in figure~\ref{fig_models}a).
We utilize the ReLU function~\cite{NH2010} for~$\varphi$, which is known to be effective for addressing the vanishing gradient problems in deep neural networks.
For determining the weights $W$, the Adam algorithm~\cite{kingma2014} is utilized.
In the present model training, early stopping~\cite{prechelt1998} with $20$ training epochs is also applied to avoid overfitting the machine-learning model.

\subsection{Convolutional neural network}
\label{sec:cnn}

Since the full flow field estimation requires a large number of spatial grid points (high spatial degrees of freedom), the computational burden is substantial for the direct application of MLP to the full flow field reconstruction~\cite{MFZNF2021,Fukamipump2022}.
To address this issue, we combine MLP and a two-dimensional convolutional neural network (CNN)~\cite{LBBH1998}.
The CNN enables regression while greatly reducing computational costs through filter sharing. 
The two-dimensional convolutional operation is illustrated in figure~\ref{fig_models}b), whose internal procedure is expressed as 
\begin{equation}
    q^{(l)}_{ijg}=\varphi\left(\sum_{l=1}^F\sum_{p=0}^{H-1}\sum_{q=0}^{H-1}h^{(l)}_{pqlg}q^{(l-1)}_{i+p-C,j+q-C,l}+b_g^{(l)}\right),
    \label{eq:CNN}
\end{equation}
where $C=\lfloor H/2\rfloor$, $H$ is the width and height of the filter, $F$ is the number of input channels, $g$ is the number of output channels, $b$ is the bias, and $\varphi$ is the activation function.
The input sensor measurements ${\bm s}\in \mathbb{R}^{n_{\bm s}}$ are transformed to a high-dimensional representation $\hat{\bm q} \in \mathbb{R}^{n_{\hat{\bm q}}}$ through the MLP for the wake estimation.
This representation $\hat{\bm q} \in \mathbb{R}^{n_{\hat{\bm q}}}$ is then reshaped into a two-dimensional matrix form $\hat{\bm q} \in \mathbb{R}^{n_{\hat x} \times n_{\hat y}}$ so that the data can be managed with a two-dimensional CNN, as illustrated in figure~\ref{fig_c2}a).
Through the CNN process in equation~\ref{eq:CNN} and upsampling operation, the present model extracts the relationship between the input sensors and the vorticity field ${\omega} \in \mathbb{R}^{n_x \times n_y}$.
As with the MLP training, we apply the ReLU function~\cite{NH2010} as the nonlinear activation function, the Adam algorithm~\cite{kingma2014} for updating filters, and early stopping~\cite{prechelt1998} to prevent overfitting.

\subsection{Long short-term memory-assisted transfer learning}

To improve the present estimation, we also utilize the long short-term memory (LSTM) algorithm~\cite{HS1997}.
LSTM is one of the recurrent neural network methods, which is suitable for predicting temporal behaviors from time-series data.
Since LSTM can hold the time-series data as memory inside the function referred to as cell, the implementation of LSTM can greatly help with the present problem that is dependent on past flow states due to its transient nature.

An LSTM layer is constructed by four functions; a cell $C$, an input gate $d$, an output gate $o$, and a forget gate $g$. 
These functions play important roles in deciding how past information is incorporated to predict the output variables. The input gate $d$ determines how much of the current information from the input of cell $e_t$ is used for prediction,
\begin{align}
d_t&=\sigma(W_d\cdot [\tilde{q}_{t-1}, e_t]+\beta_d),
\end{align}
where $q$ is the output of cell, $W$ and $\beta$ represent the weights and the bias, respectively, for each gate denoted by its subscript; the subscripts $t$ and $t-1$ represent the time indices, and $\sigma$ is the sigmoid function.
Here, the concatenation of two inputs in a model is denoted as $[{m}, {n}]$. 
In parallel, the LSTM also considers how much of the past information is kept from the cell state at the previous cell state $C_{t-1}$ using the forget gate~$g$,
\begin{align}
g_t&=\sigma(W_g\cdot [\tilde{q}_{t-1}, e_t]+\beta_g).
\end{align}
With the temporal cell state at the current time step,
\begin{align}
\widetilde{C}_{t}&=\tanh({W_c\cdot [\tilde{q}_{t-1}, e_t]+\beta_c}),
\end{align}
and the previous cell state $C_{t-1}$, the current cell state $C_t$ is determined by balancing the input gate $d$ and the forget gate $g$,
\begin{align}
C_t&=g_t C_{t-1}+d_t\widetilde{C}_t.
\end{align}
Note that the sigmoid functions used for the input and the output gates play important roles in avoiding gradient vanishing problems.
At the output of the LSTM layer, the amount of information at the cell state $C_t$ being leveraged for short-term prediction (i.e., the output at the next step ${\tilde{q}}_t$) is assessed using the output gate $o$ with
\begin{align}
o_t&=\sigma(W_o\cdot [\tilde{q}_{t-1}, e_t]+\beta_o),\\
{\tilde{q}}_t&=o_t\tanh(\it{C_t}).
\end{align}
With this formulation, the LSTM is able to predict the variable at the next step ${\tilde{q}}_t$ while considering the long-term memory influence with the concept of cell state $C$.

\begin{figure}
  \centering
    \includegraphics[width=\textwidth]{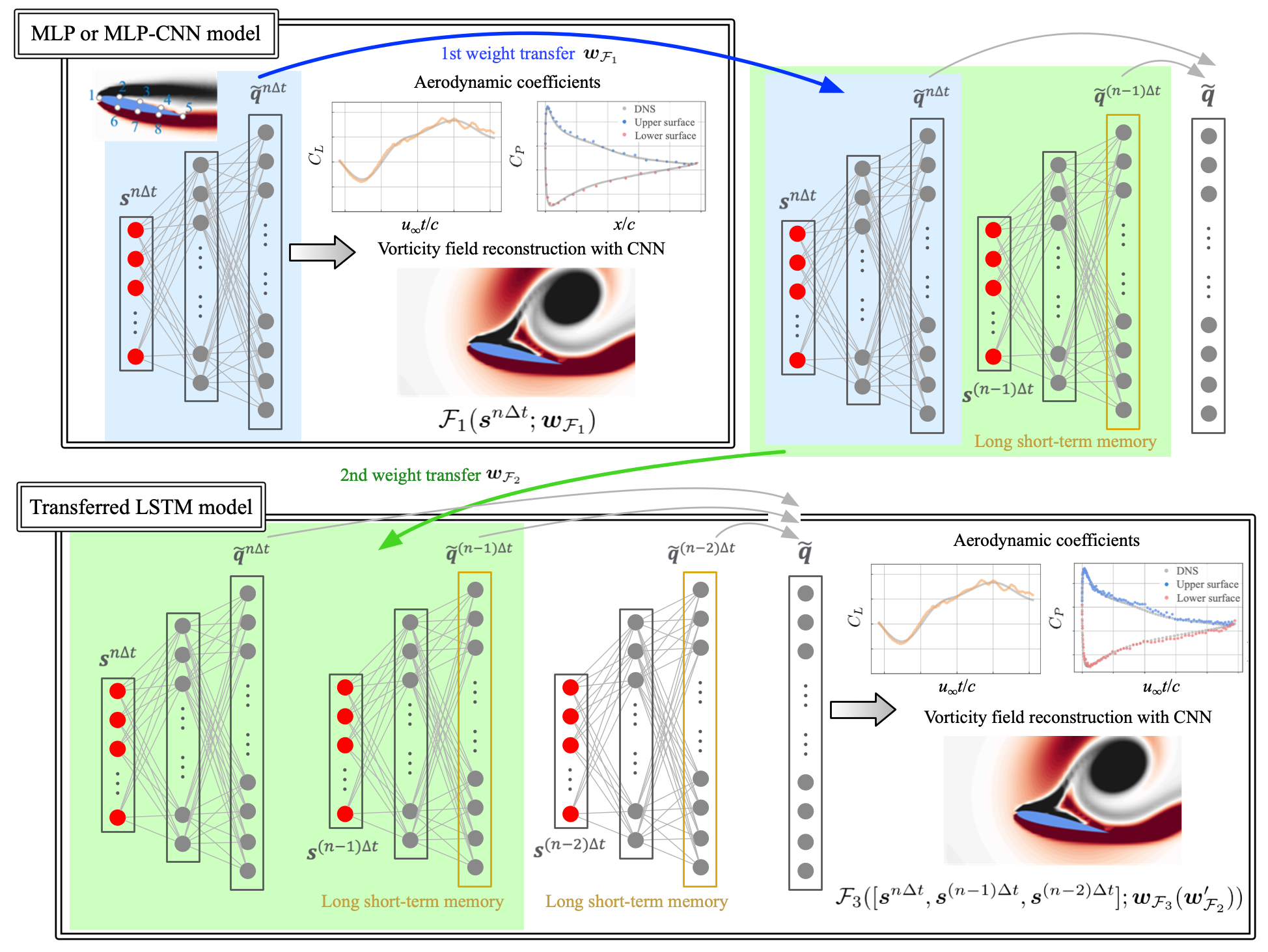}
    \caption{Long short-term memory-assisted transfer learning.
    }
    \label{fig_c22}
\end{figure}

Here, we combine the high-dimensional representation of the input measurements obtained through the MLP $\tilde{\bm{q}}^{n\Delta t}$ with two previous time sequences extracted by LSTMs $\{\tilde{\bm{q}}^{(n-1)\Delta t},\tilde{\bm{q}}^{(n-2)\Delta t}\}$ such that ${\tilde{\bm{q}}=[\tilde{\bm{q}}^{n\Delta t}+\tilde{\bm{q}}^{(n-1)\Delta t}+\tilde{\bm{q}}^{(n-2)\Delta t}}]$, as illustrated in figure~\ref{fig_c22}.
This combined vector $\tilde{\bm{q}}$ with three time steps is then provided to the MLP layer of the force estimation and the $C_p$ estimation, or the two-dimensional CNN layer of the vorticity reconstruction task.

Moreover, we utilize the concept of transfer learning for the LSTM-assisted network.
Transfer learning can facilitate the training process by setting appropriate initial weights~\cite{pan2009survey}.
The present strategy of the LSTM-assisted transfer learning is graphically summarized in figure~\ref{fig_c22}.
In the present study, the weights of pre-trained MLP ${\bm w}_{\cal M}$ are adopted as initial weights of the second model ${\cal F}_2$ which has two sensor input gates ${\bm s}^{(n-1)\Delta t}$ and ${\bm s}^{n\Delta t}$.
The high-dimensional feature of input sensor measurements $\hat{\bm q}$ from the MLP part of the model is merged with that from LSTM.
Once the training for the second model ${\cal F}_2$ is completed, the optimized weights of the second model ${\bm w}_{{\cal F}_2}$ are repeatedly transferred to the third model ${\cal F}_3$ which considers sensor measurements at three different time steps ${\bm s}^{(n-2)\Delta t}$, ${\bm s}^{(n-1)\Delta t}$, and ${\bm s}^{n\Delta t}$.
The weight optimizations through these operations are mathematically expressed as
\begin{align}
    {\bm w}_{{\cal F}_1} &= {\rm argmin}_{{\bm w}_{{\cal F}_1}}||{\bm q}-{{\cal F}_1}({\bm s}^{n\Delta t};{\bm w})||_2,\\
    {\bm w}_{{\cal F}_2} &= {\rm argmin}_{{\bm w}_{{\cal F}_2}}||{\bm q}-{\cal F}_2([{\bm s}^{n\Delta t},{\bm s}^{(n-1)\Delta t}]; {{\bm w}_{{\cal F}_2}}({\bm w}'_{{\cal F}_1}))||_2,\\
    {\bm w}_{{\cal F}_3} &= {\rm argmin}_{{\bm w}_{{\cal F}_3}}||{\bm q}-{\cal F}_3([{\bm s}^{n\Delta t},{\bm s}^{(n-1)\Delta t},{\bm s}^{(n-2)\Delta t}]; {\bm w}_{{\cal F}_3}({\bm w}'_{{\cal F}_2}))||_2,
\end{align}
where ${\bm w}'_{{\cal F}_1}$ denotes the weights assigned to the common part of the first MLP-CNN model and the second model, ${\bm w}'_{{\cal F}_2}$ represents the weights assigned to the common part of the second MLP-LSTM-CNN model and the third model, respectively.
Since transfer learning can aid in the computational reduction by enabling fast convergence of weights~\cite{MFZF2020,guastoni2020convolutional}, we can expect accurate flow reconstruction with minimal training costs using transfer learning with LSTM.

\section{Results and Discussions}
\label{sec:res}

\subsection{Aerodynamic forces}
\label{sec:cdcl}

\begin{figure}
  \centering
    \includegraphics[width=\textwidth]{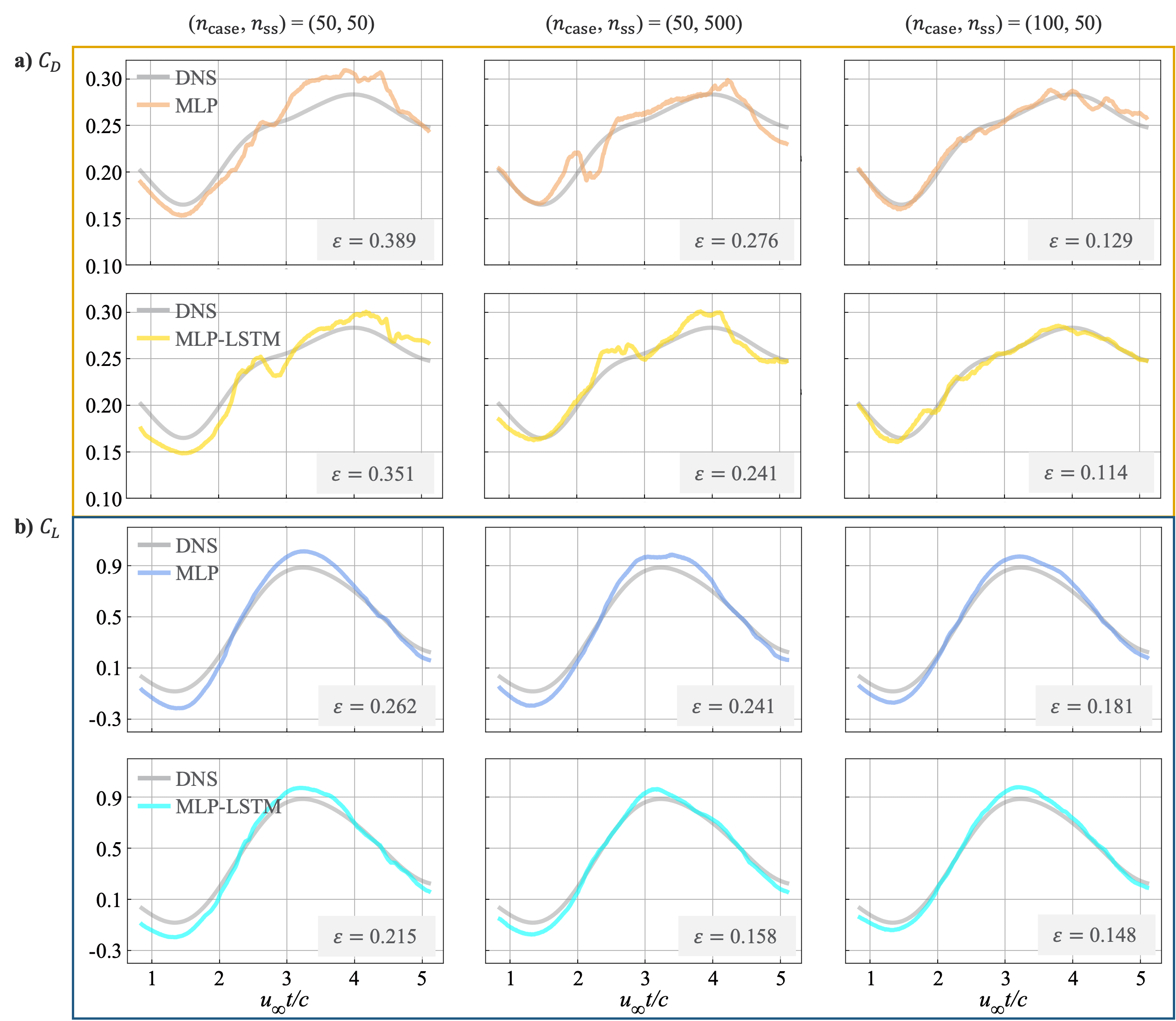}
    \caption{Estimation of $C_D$ and $C_L$ obtained with MLP and MLP-LSTM models for the case of $(u_{\theta \rm{max}}/u_{\infty}, R/c, y_{0}/c)=(-0.19, 0.83, 0.28)$.
    }
    \label{fig_CD}
\end{figure}

{Let us first present the machine-learning-based estimation of $C_L$ and $C_D$ from the pressure sensor inputs. 
Here, we here prepare machine learning models ${\cal F}$ for each coefficient such that ${C_L}={\cal F}_L({\bm s}(t))$ and ${C_D}={\cal F}_D({\bm s}(t))$.
The estimation results for $C_D$ and $C_L$ are shown in figure~\ref{fig_CD}. When training with only 50 training cases with each case having 50 snapshots, the model achieves a qualitative estimation of $C_D$. 
Here, we denote the number of cases as $n_{\rm case}$, and the number of snapshots per case as $n_{\rm ss}$.
We also quote the $L_2$ error norm $\varepsilon\equiv||{\bm f}_{\rm Ref}-{\bm f}_{\rm ML}||_2/||{\bm f}_{\rm Ref}-\overline{\bm f}||_2$, where ${\bm f}_{\rm Ref}$ and ${\bm f}_{\rm ML}$ are the reference and the machine-learning-based estimation, respectively, of variable ${\bm f}$.
Note that this error is normalized by the fluctuation of a variable from its steady state value $\overline{\bm f}$. For $C_L$ and $C_D$, the error is measured over the time range $u_{\infty}t/c = [0.85,5.1]$ for each case.

The estimation results show that the positions of the peak and trough of $C_D$ induced by the vortex-airfoil wake interaction are qualitatively predicted, yet the exact values are off from the DNS result. 
Increasing the number of training cases $n_{\rm{case}}$ improves the estimation performance. Enhanced agreement between the estimation and DNS is also achieved when increasing the number of snapshots to 500, as illustrated in figure~\ref{fig_CD}a). The enhancement in the data diversity leads to a drastic decrease in the prediction error.
In contrast to  50 training cases, utilizing 100 training cases yields a $67\%$ deduction in test error. The reason why the expansion of training cases is beneficial for prediction performance is that the machine learning model can cover a larger parameter space, which assists in better predicting unseen test cases. Yet, considering the vast parameter space, 100 training cases are very few.

To obtain an accurate reconstruction while using as little data as possible, we then incorporate the transfer-learning-based LSTM into the model for $C_D$, as shown in figure~\ref{fig_CD}a). 
Due to the transient nature of the current vortex-airfoil interaction problem, the present transfer-learning-based LSTM is able to build a reliable connection between sensor input and output based on historical information. 
For all three examples, using the MLP-LSTM model gives rise to a 10\% decrease in the estimation error.

Estimation for $C_L$ is presented in figure~\ref{fig_CD}b). Enhancement in the reconstruction of $C_L$ from increasing the amount of training data is also shown in figure~\ref{fig_CD}b). The new MLP-LSTM model reduces the test error to 0.215, 0.158, and 0.148 for $(n_{\rm case}, n_{\rm ss})=(50,50)$, $(n_{\rm case}, n_{\rm ss})=(50,500)$ and $(n_{\rm case}, n_{\rm ss})=(100,50)$, respectively. Similar to the $C_D$ estimation, the transfer-learned-LSTM architecture is also useful in the estimation of $C_L$. We also note that the reconstruction for $C_L$ is usually better than $C_D$, which is due to the variation of $C_L$ over time being much larger than that of $C_D$.

}

\subsection{Surface pressure distribution}
\label{sec:pres}

\begin{figure}[t]
  \centering
    \includegraphics[width=\textwidth]{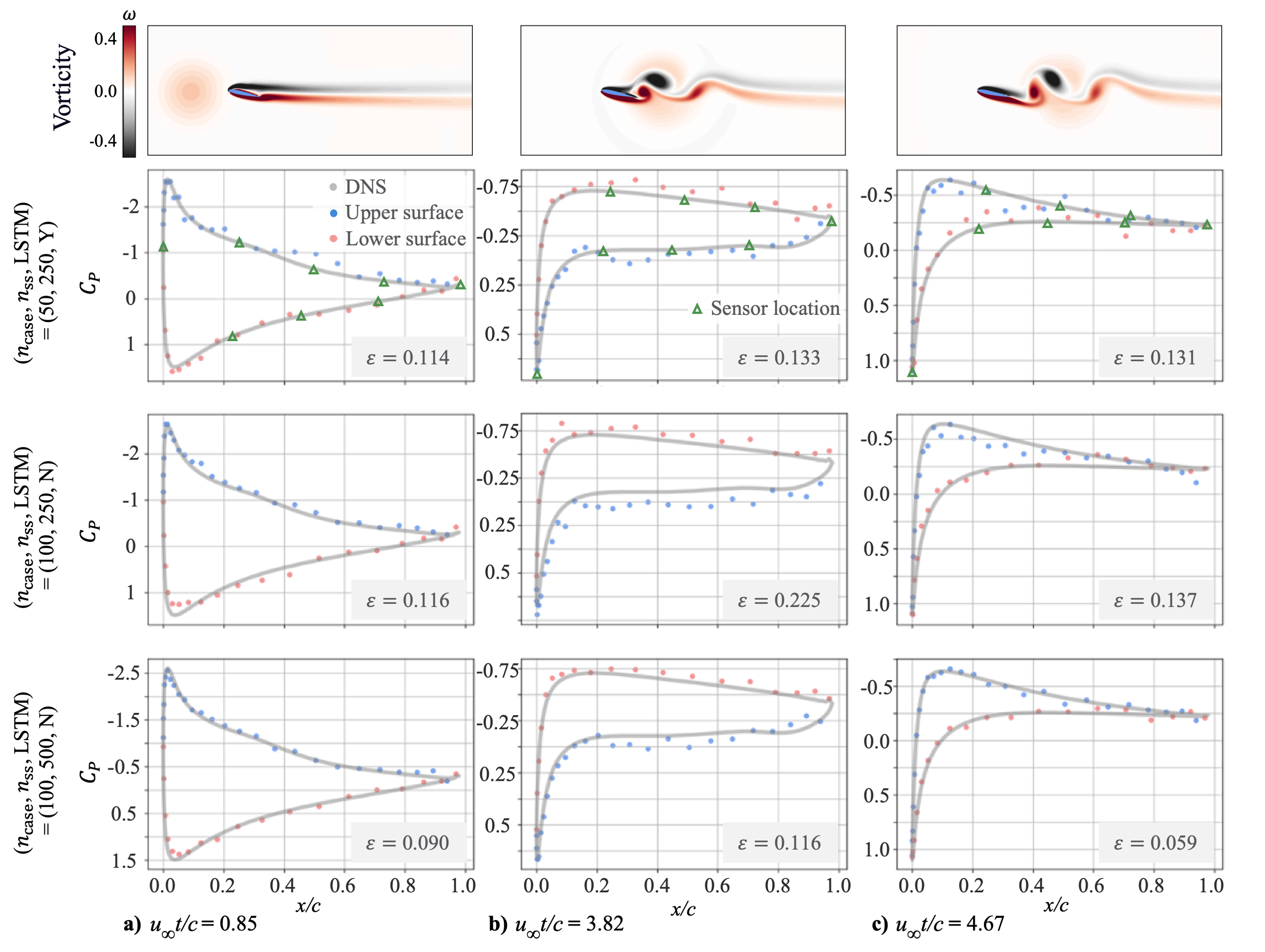}
    \caption{Machine-learning-based estimation of the pressure distribution over an airfoil surface.
    The pressure coefficient $C_p$ at $u_{\infty}t/c=$ a)$~0.85$, b)$~3.82$, and c)$~4.67$. Results are shown for the case of: $(u_{\theta \rm{max}}/u_{\infty}, R/c, y_{0}/c)=(0.35, 0.95, -0.15)$.
    } 
    \label{fig_r2}
\end{figure}

Next, we perform the MLP-based estimation of the pressure distribution $C_p$ over the airfoil surfaces.
Representative snapshots of the vorticity field for a test case when a vortical disturbance passes around the airfoil are shown in the first row of figures~\ref{fig_r2}.
Similar to the aerodynamics forces $C_L$ and $C_D$, the reconstruction performance is strongly influenced by the number of cases $n_{\rm case}$, the number of snapshots per case $n_{\rm ss}$, as well as whether transfer-learned LSTM is incorporated. 
When training the MLP-LSTM model with 50 cases and 250 snapshots per case, a qualitative reconstruction is achieved for $C_p$. 
As shown in the first row of figure~\ref{fig_r2}, the estimated $C_p$ at both the upper and lower surfaces of the airfoil are in agreement with the DNS.
As we increase the number of cases from 50 to 100 without utilizing transfer-learned LSTM, this machine-learning model also reconstructs $C_p$ in a reasonable manner, as shown in the second row of figure~\ref{fig_r2}. 
However, by comparing the reconstruction of $C_p$ of $(n_{\rm case}, n_{\rm ss})=(100,250)$  without LSTM against the results with LSTM implemented, it is found that the use of transfer-learned LSTM greatly improves reconstruction, achieving enhanced performance with only half of the training data.
In order to further improve the estimation performance, increasing the number of snapshots from 250 to 500 for 100 training cases achieves a similar performance as the results of $(n_{\rm case}, n_{\rm ss},\rm LSTM)=(50,250,\rm{Y})$, as shown in the last row of figure~\ref{fig_r2}.
Note that although we are using 100 training cases and 500 snapshots per case, the training data is still small compared to the broad parameter space of the test cases.

\subsection{Vorticity field}
\label{sec:2D}

\begin{figure}
  \centering
    \includegraphics[width=\textwidth]{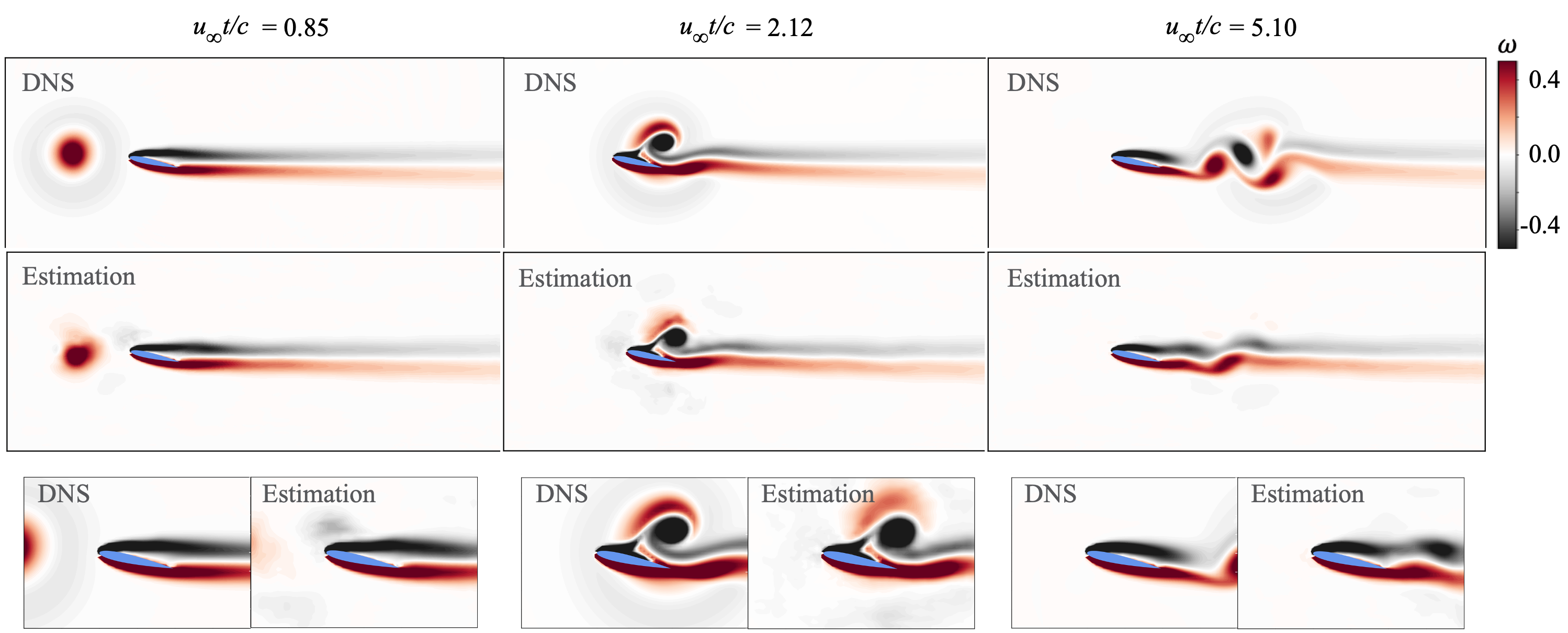}
    \caption{Reconstructed vorticity flow field with large region training and windowed region training. Results shown for $(u_{\theta \rm{max}}/u_{\infty}, R/c, y_{0}/c)=(0.65, 0.40, 0)$.
    }
    \label{fig_r4}
\end{figure}

{

We employ the present machine learning techniques to reconstruct the two-dimensional vorticity field from sensor measurements on an airfoil using the MLP-CNN model.
Analogous to the results in section~\ref{sec:cnn}, the combination of MLP and CNN is suitable to estimate the vortical flow from sensors.
The reconstruction of the spatially discretized vorticity field ${\omega}\in \mathbb{R}^{100\times 200}$ is summarized in figure~\ref{fig_r4}.
The present model successfully captures the vortical disturbance at $u_{\infty}t/c=0.85$. 
The location and the strength of the vortex are well reconstructed. 
The interaction between the vortex disturbance and the flow field around the airfoil at $u_{\infty}t/c=2.12$ is also reproduced well. This is approximately the time at which $C_L$ and $C_D$ drop to their minimum values, serving as an important dynamic transition point. 
However, the wakes behind the trailing edge at $u_{\infty}t/c=5.10$ are not accurately reconstructed because these wake structures are farther away from the airfoil during this period. 
Sensors on the airfoil surface measure do not observe a sizeable change in pressure, making it difficult to reconstruct far-field wakes, which is expected.

Let us now focus on the critical near-wake region around an airfoil since this region primarily determines the unsteady loading.
Considering only the near-field region enables us to greatly reduce the size of training data and the associated computational costs.
Results from training with a smaller region are described in figure~\ref{fig_r4}.
The windowed training model also provides improved estimations with lower error.
The averaged $L_2$ error for the test case reduces from 0.329 (large region training) to 0.261 (windowed region training), which also shows the influence of the region size to estimate the wake field with the modest computational cost.

\begin{figure}[t]
  \centering
    \includegraphics[width=0.9\textwidth]{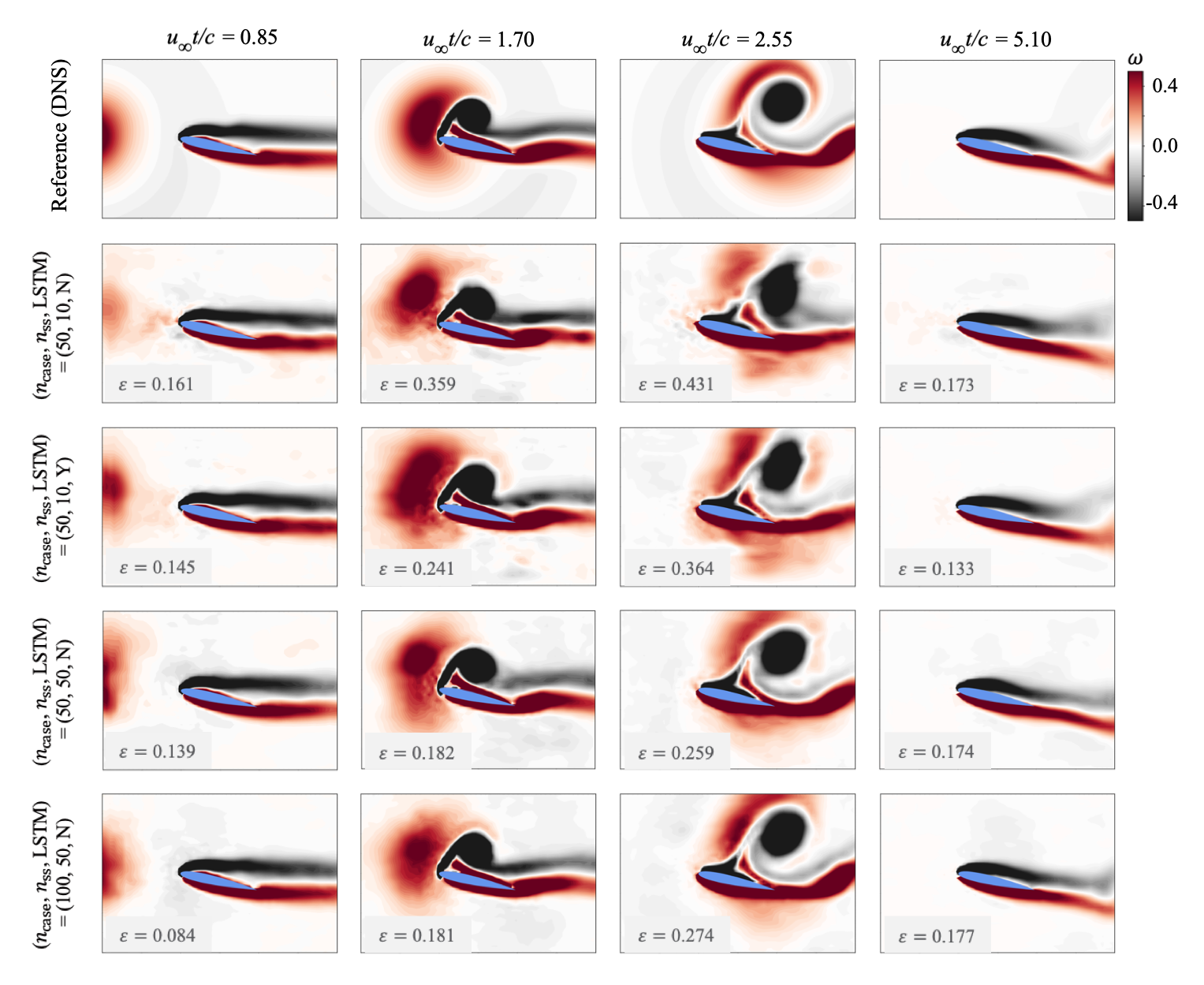}
    \caption{Dependence of the reconstruction accuracy on the present enhancement methods with window training for vorticity wake problem. 
    Results are shown for the case $(u_{\theta \rm{max}}/u_{\infty}, R/c, y_{0}/c)=(0.72, 0.64, -0.10)$.
    }
    \label{fig_r5}
\end{figure}

Moreover, the enhancement in reconstructing the vorticity field can also be achieved by increasing the amount of training data and utilizing transfer-learned LSTM, as summarized in figure~\ref{fig_r5}. 
For all time series, a qualitative and insightful reconstruction of the vorticity field is achieved with as less as 10 snapshots per case, as shown in the second column in figure~\ref{fig_r5}. 
When we apply the transfer-learned-LSTM to the same dataset, up to 33\% reduction in $L_2$ error is accomplished. 
Additionally, increasing the number of snapshots to 50 or increasing the data diversity by using 100 training cases produces further improvements.

It is worth noting that the reconstruction accuracy for the interaction process is not uniform.
For example, at $u_{\infty}t/c=1.70$ and $u_{\infty}t/c=2.55$ when the center of the disturbance is near the airfoil, the $L_2$ errors are relatively high. Due to the high level of interaction, the complex morphological changes in the vorticity field result in an increased error.
However, this does not indicate that the machine-learning model is not able to extract the crucial features of the flow field. Instead, the errors are partially due to the modest displacement of vortical structures. 
In addition, the reconstructed vorticity field at $u_{\infty}t/c=5.10$ shows that the transfer-learned LSTM shows its superiority in estimating the small fluctuation behind the trailing edge compared to the enhancement of data amount or diversity. 
Based on the insights gained from this study, we deduce that when the influence from the disturbance is greater (strong disturbances with large sizes and the interactions around the airfoil), the accuracy of the reconstruction is improved. Here again, the transfer-learned LSTM greatly improves the estimation for the overall dynamic process.}

\subsection{Influence on the sensor positions}

\begin{figure}[t]
  \centering
    \includegraphics[width=0.65\textwidth]{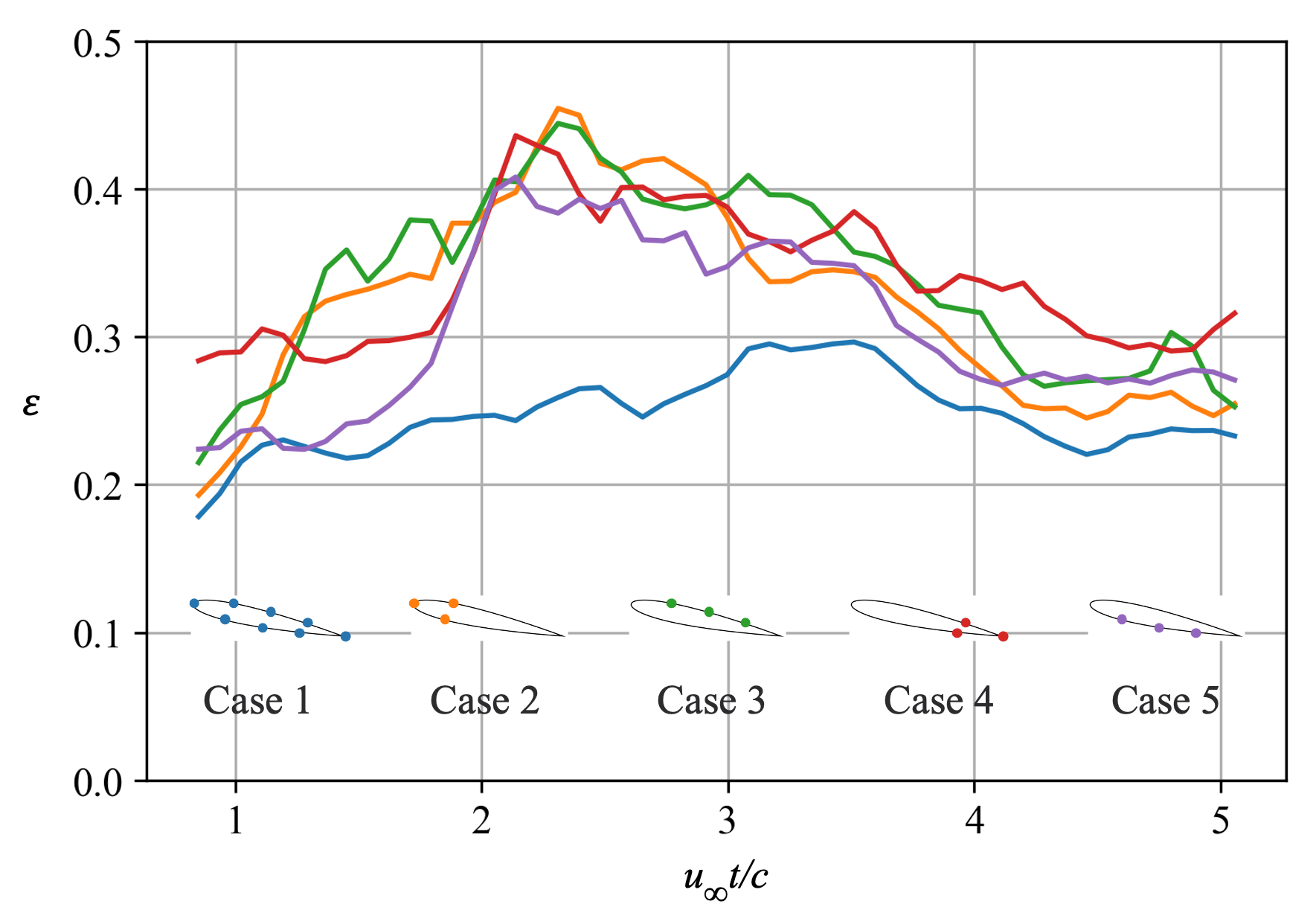}
    \caption{
    {Dependence of the $L_2$ errors on the sensor positions. 
    Cases 1 to 5 denote 8 uniform sensors, 3 leading edge sensors, 3 top surface sensors, 3 trailing edge sensors, and 3 bottom surface sensors, respectively.
    The machine-learning model with the condition of $(n_{\rm case},n_{\rm sss},\rm {LSTM})=(50,50,Y)$ is used.}
    }
    \label{fig_sensor_error}
\end{figure}

\begin{figure}[t]
  \centering
    \includegraphics[width=1\textwidth]{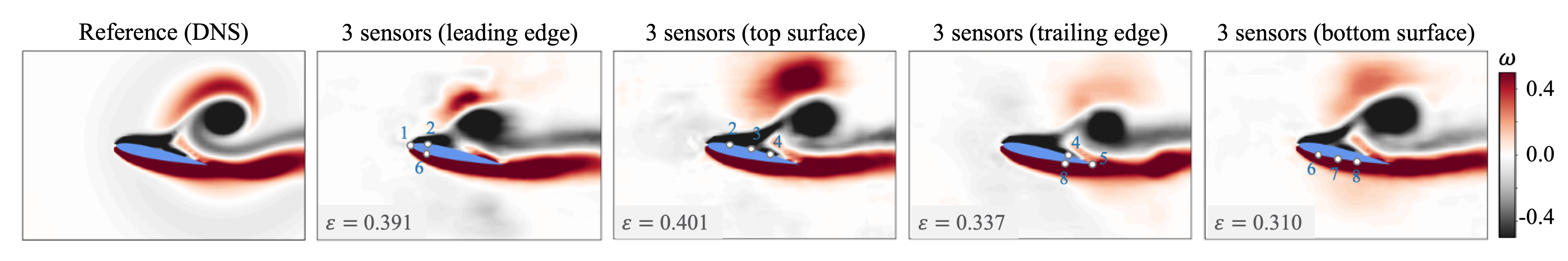}
    \caption{Dependence of the reconstruction accuracy on the location of sensors for the reconstructed vorticity wake. $(n_{\rm case},n_{\rm sss},\rm {LSTM})=(50,50,Y)$
    As a test case, we use $(u_{\theta \rm{max}}/u_{\infty}, R/c, y_{0}/c)=(0.65, 0.40, 0)$.
    }
    \label{fig_r7}
\end{figure}

Next, let us examine the estimation performance of the machine-learning models trained with different numbers and placements of the sensors.
As shown in figure~\ref{fig_sensor_error}, we consider the uses of 8 sensors (case 1), 3 sensors around the leading edge (case 2), 3 sensors on the top surface (case 3), 3 sensors around the trailing edge (case 4), and 3 sensors on the bottom surface (case 5), respectively. 
With 8 sensors (case 1), the lowest $L_2$ error is achieved compared to the other cases with 3 sensors, as expected.
With 3 sensors, we observe that Case 5 with the bottom surface sensors usually presents a lower error than Cases 2 to 4 for the whole time range.
This is likely because the sensors on the pressure side may sense the vortical structures approaching an airfoil easier and earlier than having sensors on the suction side.

We also assess the estimation performance over time in figure~\ref{fig_sensor_error}. 
Before the vortical disturbance impinges on the airfoil ($u_{\infty}t/c<2$) and after the vortex moves away from the trailing edge of the airfoil ($u_{\infty}t/c>4$), we observe relatively low $L_2$ error.
For $u_{\infty}t/c \in [2,4]$, due to the complex interactions between the disturbance and the airfoil, the estimation for this time period is more difficult than other times.
However, we note that the present model still achieves qualitative reconstructions even for the strong vortex-airfoil wake interaction process, as depicted in figure~\ref{fig_r7}.
These reconstructed snapshots correspond to the moment $u_{\infty}t/c=2.12$ for $(u_{\theta \rm{max}}/u_{\infty}, R/c, y_{0}/c)=(0.65, 0.40, 0)$. 
This implies that monitoring not only the scalar error measurement but also the reconstructed flow fields is essential for appropriate assessments of machine-learning-based flow estimations. These results also provide practical insights into the choice of sensor locations.
It is recommended that sensors are placed on the suction and pressure sides for the present problem.

\subsection{Robustness against noisy sensor measurements}
\label{sec:noise}

\begin{figure}
  \centering
    \includegraphics[width=1\textwidth]{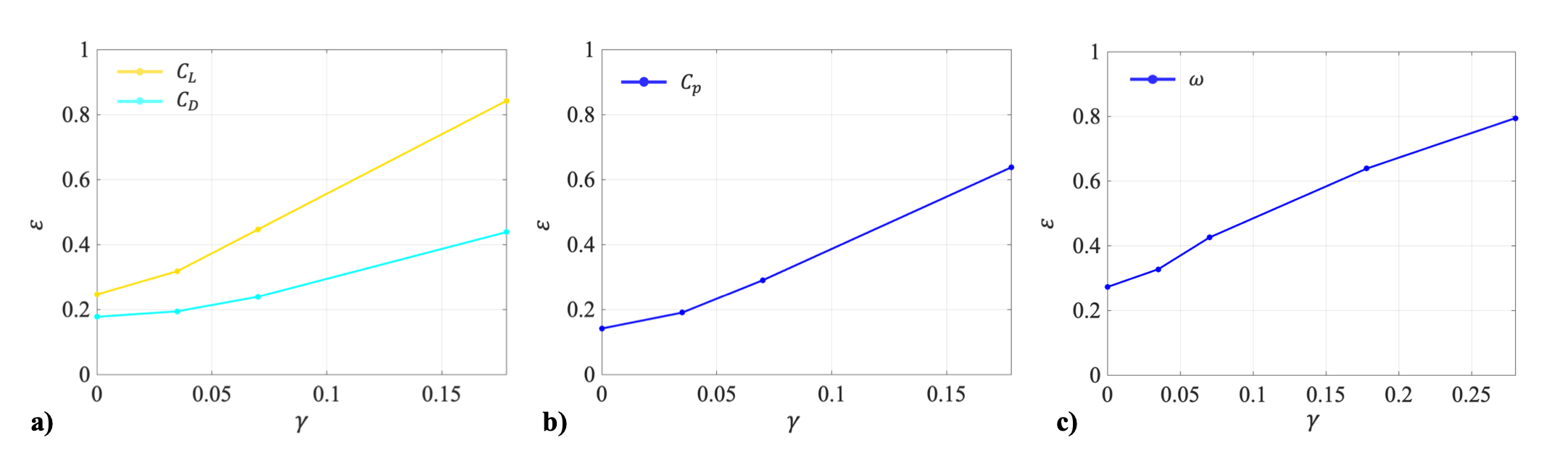}
    \caption{Machine-learning model robustness against noisy sensor measurements. a)~ $C_L$ and $C_D$, $(n_{\rm case},n_{\rm ss},\rm {LSTM})=(100,50,Y)$, b)~$C_P$, $(n_{\rm case},n_{\rm ss},\rm {LSTM})=(100,500,Y)$, and c)~Vorticity field $\omega$, $(n_{\rm case},n_{\rm ss},\rm {LSTM})=(100,100,Y)$.}
    \label{noise_CLCD}
\end{figure}

\begin{figure}
  \centering
    \includegraphics[width=1\textwidth]{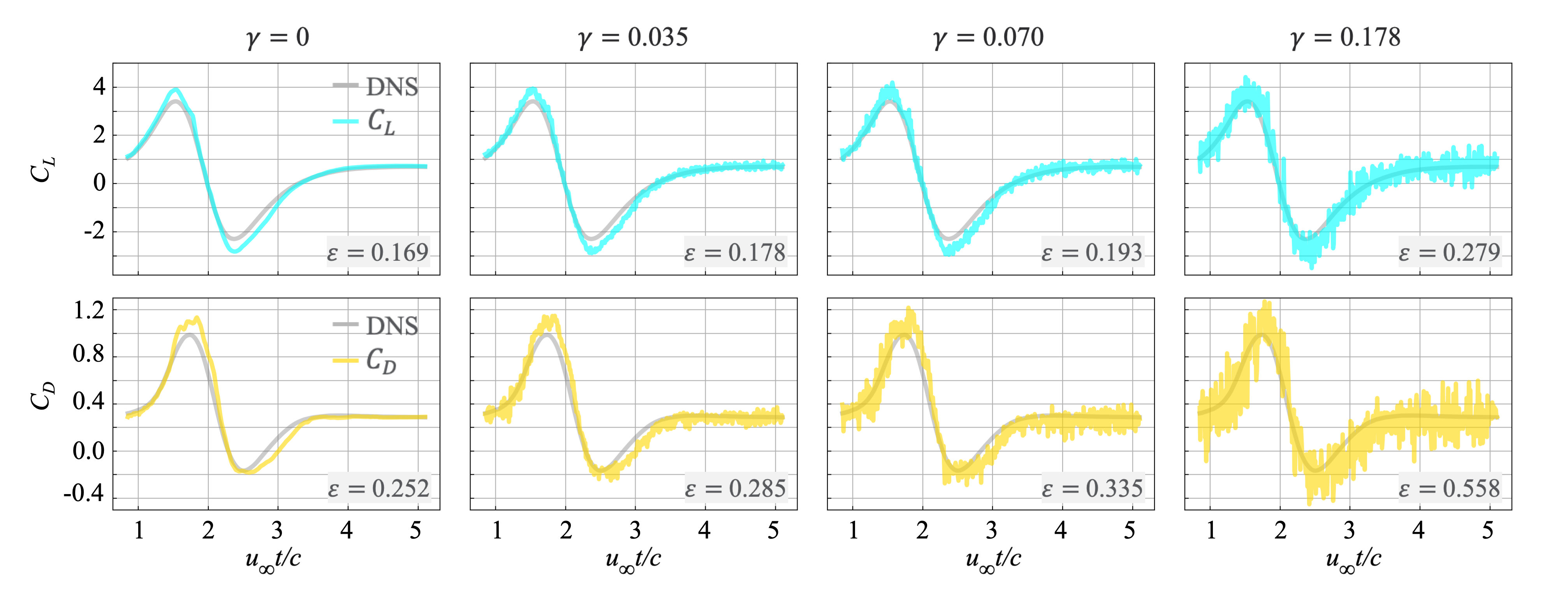}
    \caption{Reconstruction of $C_L$ and $C_D$ subjected to different levels of input noise. Results are shown for $(u_{\theta \rm{max}}/u_{\infty}, R/c, y_{0}/c)=(0.96, 0.57, 0.21)$.
    }
    \label{noise_compare_CLCD}
\end{figure}

\begin{figure}
  \centering
    \includegraphics[width=1\textwidth]{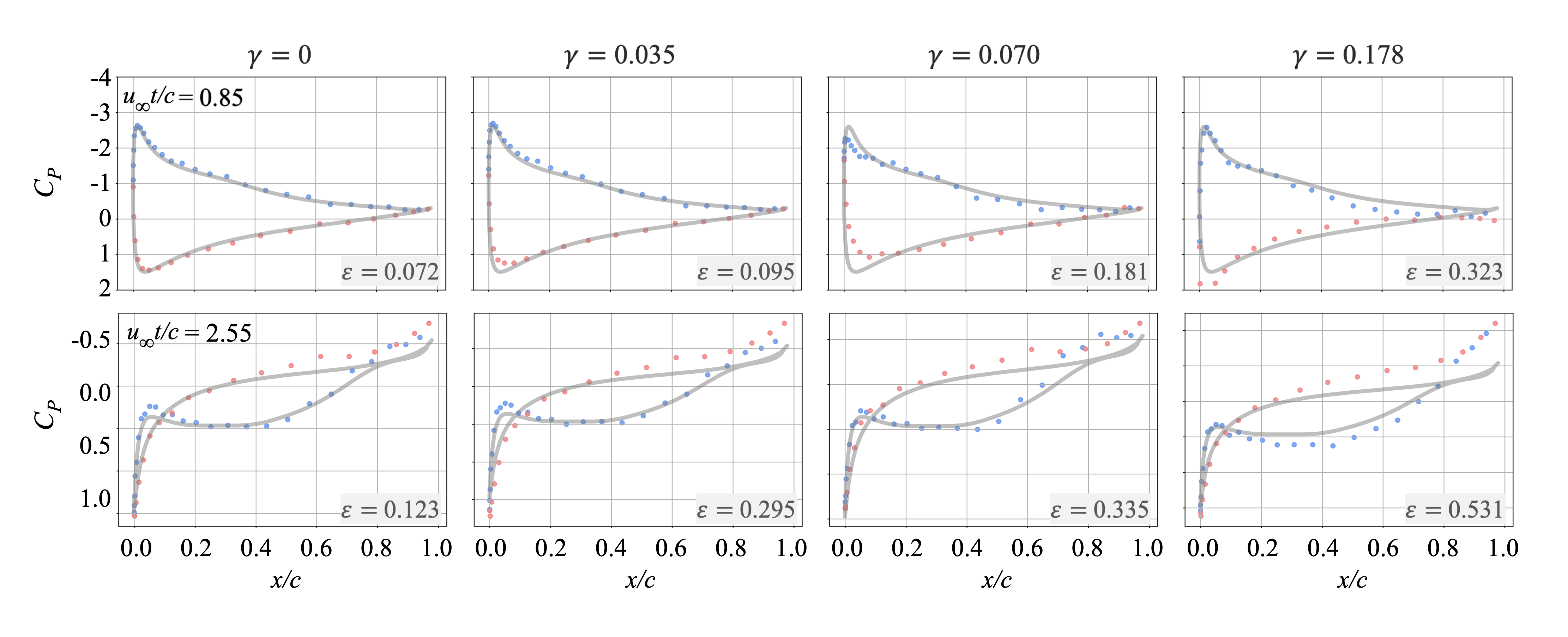}
    \caption{Comparison of $C_P$ subjected to different levels of input noise. Results are shown for $(u_{\theta \rm{max}}/u_{\infty}, R/c, y_{0}/c)=(0.35, 0.93, -0.15)$.
    }
    \label{noise_compare_CP}
\end{figure}

\begin{figure}
  \centering
    \includegraphics[width=1\textwidth]{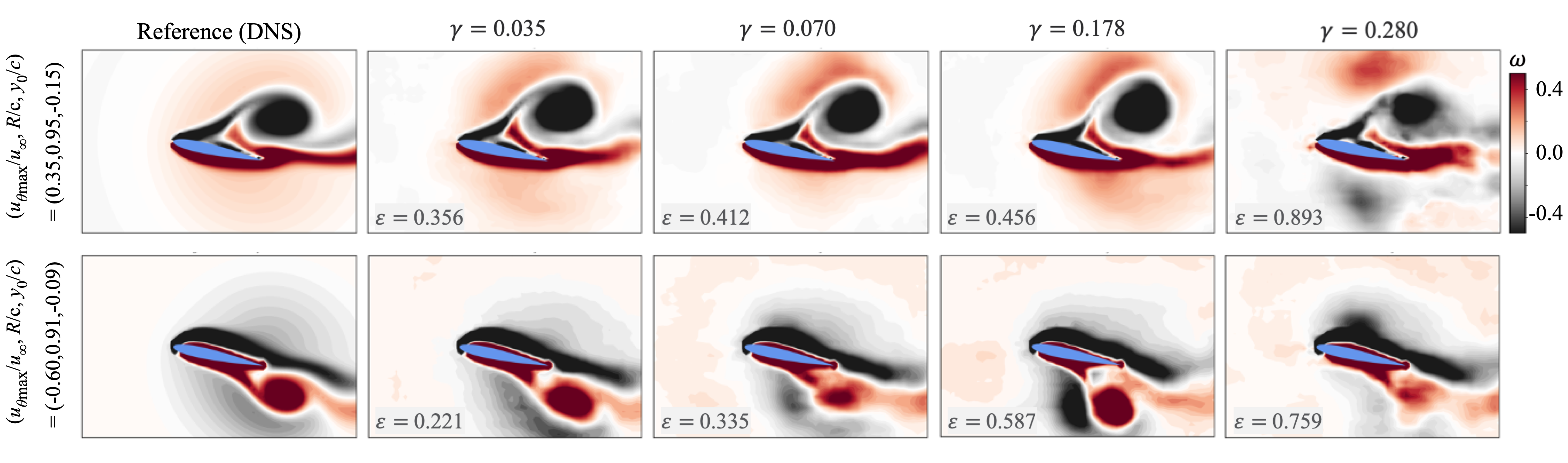}
    \caption{Comparison of estimated vorticity fields subject to different levels of input noise.
    }
    \label{noise_compare_vort}
\end{figure}

Let us evaluate the machine-learning model robustness against the noisy sensor measurements.
We use the Gaussian noise $\bm n$ for the sensor input $\bm s$.
Hence, the estimated output is expressed as
\begin{equation}
    {\bm q}_{n} = {\cal F}({\bm s}+ \bm n)
\end{equation}
where ${\bm q}_{n}$ is the output of the model, ${\cal F}$ is the model trained without noisy inputs, and $\gamma = \Vert {\bm n}\Vert/\Vert {\bm s}\Vert$ is the magnitude of the noise.

The estimation performance of $C_L$, $C_D$, $C_P$, and vorticity field with noisy inputs (pressure measurements) are considered herein, as shown in figure~\ref{noise_CLCD}. 
For all estimations, the error increases with the magnitude of the input noise, as expected. 
The reconstructed $C_L$, $C_D$, $C_P$, and vorticity field are also shown in figures~\ref{noise_compare_CLCD}-\ref{noise_compare_vort}.
Regarding the estimated $C_L$ and $C_D$ in figure~\ref{noise_compare_CLCD}, the reconstructed $C_L$ and $C_D$ present smooth curves without noisy input of $\gamma = 0$.
With increasing $\gamma$, $C_L$ and $C_D$ have high fluctuations resulting in a larger $L_2$ error but with the overall trend well reproduced.

The estimated $C_P$ from noisy pressure inputs is depicted in figure~\ref{noise_compare_CP}.
While the error solely increases with the noise magnitude, we find that the error at $u_{\infty}t/c=2.55$ is larger than that at $u_{\infty}t/c=0.85$.
This is caused by the intense wake-vortex gust interaction at $u_{\infty}t/c=2.55$ which induces rapid changes in the pressure distribution on the airfoil surface.
Although the error reports approximately 0.5 with $\gamma=0.178$, the whole trend of the $C_P$ curve is well-estimated, supporting the robustness of the present machine-learning model.

The reconstructed vorticity fields across the different levels of noisy inputs are also exhibited in figure~\ref{noise_compare_vort}. 
We show two cases of the vortical disturbance, $(u_{\theta \rm{max}}/u_{\infty}, R/c, y_{0}/c)=(0.35, 0.95, -0.15)$ and $(-0.60, 0.91, -0.09)$.
A large positive disturbance is introduced in the former case, while a negative vortical gust travels over the airfoil in the latter case.
In the case of $(u_{\theta \rm{max}}/u_{\infty}, R/c, y_{0}/c)=(0.35, 0.95, -0.15)$, the estimated vorticity field retains the primary vortical features for $\gamma \le 0.178$. The estimated flow field deviates from the reference DNS field at $\gamma =0.28$.
For the case of $(u_{\theta \rm{max}}/u_{\infty}, R/c, y_{0}/c)=(-0.60, 0.91, -0.09)$, spurious negative structure attached to the trailing edge vortex emerges beyond $\gamma =0.178$, albeit the overall flow is reconstructed well. At $\gamma =0.28$, although the $L_2$ error norm is relatively high, the main wake structures are nonetheless reconstructed.
These results suggest that the present machine-learning models that incorporate dynamics are robust against noisy pressure measurements even with a small amount of training data.

\section{Concluding remarks}
\label{sec:conc}

High-fidelity machine-learning-based reconstructions are developed for aerodynamic force coefficients, pressure distribution over the airfoil, and two-dimensional vorticity flows that experience an impact with a disturbance vortex.
Such reconstruction using sparse sensor measurements and a modest amount of training data is extremely challenging due to the strong nonlinearities and the transient nature of flow fields which requires a vast parameter space to be covered during the learning process.
For accurate reconstruction, we developed machine learning models that are suitable for estimating the transient flow features.
A multi-layer perceptron is chosen for its ability in constructing the nonlinear relation between limited sensor measurements and aerodynamic forces coefficient as well as pressure over the airfoil surface. 
A convolutional neural network coupled with MLP addresses the problem of estimating the vorticity fields with rich information in an efficient way with the filtering process. To better capture dynamical features in time, long short-term memory (LSTM)-assisted transfer learning is utilized via passing information from the historical scenarios, which is embedded in the aforementioned two model structures.
Due to the transient nature of the vortex-airfoil interaction problem, the use of LSTM greatly assists in the improvement of the estimation with as few as 10 training snapshots.

The main contribution of the present study is how time-varying flows with a vast parameter space are reconstructed accurately. 
For this study, the parameter space is comprised of maximum rotational velocity, radius, and position of the disturbance vortex. 
As shown in this paper, careful sampling of training data and incorporation of dynamics into the machine-learning model is important. Based on our study, we also showed that accurate reconstruction of vortical structures is easier to accomplish for high-intensity interaction processes between the vortical disturbance and the airfoil (strong vortex with large size, interacting close to the airfoil). In addition, we accessed proper sensor locations over different time periods.
We expect that the present machine-learning-based reconstruction method will be useful in predicting and controlling flows associated with vortex-airfoil interactions in the future.

\section*{Acknowledgments}
YZ, KF and KT acknowledge Ebara Corporation for supporting this research. 
We are thankful to Akira Goto, Motohiko Nohmi, Masashi Obuchi, and Hiroyoshi Watanabe for enlightening discussions.


\bibliographystyle{unsrt}  
\bibliography{refs}

\begin{thebibliography}{10}

\bibitem{jones2022physics}
A.~R. Jones, O.~Cetiner, and M.~J. Smith.
\newblock Physics and modeling of large flow disturbances: Discrete gust
  encounters for modern air vehicles.
\newblock {\em Annu. Rev. Fluid Mech.}, 54:469--493, 2022.

\bibitem{Jeff2021}
J.~Eldredge, M.~Le Provost, R.~Baptista, and Y.~Marzouk.
\newblock Applications of ensemble {K}alman filtered vortex modeling to
  gust--wing interactions.
\newblock {\em AIAA J.}, page 1936, 2021.

\bibitem{pfnur2019leading}
S.~Pfn{\"u}r and C.~Breitsamter.
\newblock Leading-edge vortex interactions at a generic multiple swept-wing
  aircraft configuration.
\newblock {\em J. Aircr.}, 56(6):2093--2107, 2019.

\bibitem{iannelli2019worst}
A.~Iannelli, P.~Seiler, and A.~Marcos.
\newblock Worst-case disturbances for time-varying systems with application to
  flexible aircraft.
\newblock {\em J. Guid. Control Dyn.}, 42(6):1261--1271, 2019.

\bibitem{scherl2020geometric}
I.~Scherl, B.~Strom, S.~L. Brunton, and B.~L. Polagye.
\newblock Geometric and control optimization of a two cross-flow turbine array.
\newblock {\em J. Renew. Sustain.}, 12(6):064501, 2020.

\bibitem{liu2018core}
Q.~Liu, B.~An, M.~Nohmi, M.~Obuchi, and K.~Taira.
\newblock Core-pressure alleviation for a wall-normal vortex by active flow
  control.
\newblock {\em J. Fluid Mech.}, 853:R1, 2018.

\bibitem{LiuJFE}
Q.~Liu, B.~An, M.~Nohmi, M.~Obuchi, and K.~Taira.
\newblock Active flow control of a pump-induced wall-normal vortex with steady
  blowing.
\newblock {\em J. Fluids Eng.}, 142(8):081202, 2020.

\bibitem{halder2020deep}
R.~Halder, M.~Damodaran, and B.~C. Khoo.
\newblock Deep learning based reduced order model for airfoil-gust and
  aeroelastic interaction.
\newblock {\em AIAA J.}, 58(10):4304--4321, 2020.

\bibitem{Girguis2022}
G.~Sedky, A.~Gementzopoulos, I.~Andreu-Angulo, F.~D. Lagor, and A.~R. Jones.
\newblock Physics of gust response mitigation in open-loop pitching manoeuvres.
\newblock {\em J. Fluid Mech.}, 944:A38, 2022.

\bibitem{herrmann2022gust}
B.~Herrmann, S.~L. Brunton, J.~E. Pohl, and R.~Semaan.
\newblock Gust mitigation through closed-loop control. {II.} {F}eedforward and
  feedback control.
\newblock {\em Phys. Rev. Fluids}, 7(2):024706, 2022.

\bibitem{ES1995}
R.~Everson and L.~Sirovich.
\newblock Karhunen--loeve procedure for gappy data.
\newblock {\em J. Opt. Soc. Am.}, 12(8):1657--1664, 1995.

\bibitem{BDW2004}
T.~Bui-Thanh, M.~Damodaran, and K.~Willcox.
\newblock Aerodynamic data reconstruction and inverse design using proper
  orthogonal decomposition.
\newblock {\em AIAA J.}, 42(8):1505--1516, 2004.

\bibitem{bewley2001dns}
T.~R. Bewley, P.~Moin, and R.~Temam.
\newblock {DNS}-based predictive control of turbulence: an optimal benchmark
  for feedback algorithms.
\newblock {\em J. Fluid Mech.}, 447:179--225, 2001.

\bibitem{adrian1988stochastic}
R.~J. Adrian and P.~Moin.
\newblock Stochastic estimation of organized turbulent structure: homogeneous
  shear flow.
\newblock {\em J. Fluid Mech.}, 190:531--559, 1988.

\bibitem{colburn2011state}
C.~H. Colburn, J.~B. Cessna, and T.~R. Bewley.
\newblock State estimation in wall-bounded flow systems. {Part 3}. the ensemble
  {K}alman filter.
\newblock {\em J. Fluid Mech.}, 682:289--303, 2011.

\bibitem{Steven2020}
S.~L. Brunton, B.~R. Noack, and P.~Koumoutsakos.
\newblock Machine learning for fluid mechanics.
\newblock {\em Annu. Rev. Fluid Mech.}, 52(1):477--508, 2020.

\bibitem{pawar2021physics}
S.~Pawar, O.~San, B.~Aksoylu, A.~Rasheed, and T.~Kvamsdal.
\newblock Physics guided machine learning using simplified theories.
\newblock {\em Phys. Fluids}, 33(1):011701, 2021.

\bibitem{pawar2022multi}
S.~Pawar, O.~San, P.~Vedula, A.~Rasheed, and T.~Kvamsdal.
\newblock Multi-fidelity information fusion with concatenated neural networks.
\newblock {\em Sci. Rep.}, 12(1):1--13, 2022.

\bibitem{hui2020fast}
X.~Hui, J.~Bai, H.~Wang, and Y.~Zhang.
\newblock Fast pressure distribution prediction of airfoils using deep
  learning.
\newblock {\em Aerosp. Sci. Technol.}, 105:105949, 2020.

\bibitem{erichson2020}
N.~B. Erichson, L.~Mathelin, Z.~Yao, S.~L. Brunton, M.~W. Mahoney, and J.~N.
  Kutz.
\newblock Shallow learning for fluid flow reconstruction with limited sensors.
\newblock {\em Proc. Royal Soc. A}, 476(2238):20200097, 2020.

\bibitem{fukami2021global}
K.~Fukami, R.~Maulik, N.~Ramachandra, K.~Fukagata, and K.~Taira.
\newblock Global field reconstruction from sparse sensors with {V}oronoi
  tessellation-assisted deep learning.
\newblock {\em Nat. Mach. Intell.}, 3(11):945--951, 2021.

\bibitem{FFT2019}
K.~Fukami, K.~Fukagata, and K.~Taira.
\newblock Super-resolution reconstruction of turbulent flows with machine
  learning.
\newblock {\em J. Fluid Mech.}, 870:106--120, 2019.

\bibitem{GANSRkim2021}
H.~Kim, J.~Kim, S.~Won, and C.~Lee.
\newblock Unsupervised deep learning for super-resolution reconstruction of
  turbulence.
\newblock {\em J. Fluid Mech.}, 910:A29, 2021.

\bibitem{CZXG2019}
S.~Cai, S.~Zhou, C.~Xu, and Q.~Gao.
\newblock Dense motion estimation of particle images via a convolutional neural
  network.
\newblock {\em Exp. Fluids}, 60:60--73, 2019.

\bibitem{kurtulus2015unsteady}
D.~F. Kurtulus.
\newblock On the unsteady behavior of the flow around {NACA} 0012 airfoil with
  steady external conditions at {$Re=1000$}.
\newblock {\em Int. J. Micro Air Veh.}, 7(3):301--326, 2015.

\bibitem{liu2012numerical}
Y.~Liu, K.~Li, J.~Zhang, H.~Wang, and L.~Liu.
\newblock Numerical bifurcation analysis of static stall of airfoil and dynamic
  stall under unsteady perturbation.
\newblock {\em Commun. Nonlinear Sci. Numer. Simul.}, 17(8):3427--3434, 2012.

\bibitem{di2018fluid}
G.~Di~Ilio, D.~Chiappini, S.~Ubertini, G.~Bella, and S.~Succi.
\newblock Fluid flow around {NACA} 0012 airfoil at low-{R}eynolds numbers with
  hybrid lattice {B}oltzmann method.
\newblock {\em Comput. Fluids}, 166:200--208, 2018.

\bibitem{Bres}
G.~A. Brès, F.~E. Ham, J.~W. Nichols, and S.~K. Lele.
\newblock Unstructured large-eddy simulations of supersonic jets.
\newblock {\em AIAA J.}, 55(4):1164--1184, 2017.

\bibitem{taylor1918dissipation}
G.~I. Taylor.
\newblock On the dissipation of eddies.
\newblock {\em Meteorology, Oceanography and Turbulent Flow}, pages 96--101,
  1918.

\bibitem{wu2020comprehensive}
Z.~Wu, S.~Pan, F.~Chen, G.~Long, C.~Zhang, and S.~Y. Philip.
\newblock A comprehensive survey on graph neural networks.
\newblock {\em IEEE Trans Neural Netw Learn Syst}, 32:4--24, 2020.

\bibitem{HS1997}
S.~Hochreiter and J.~Schmidhuber.
\newblock Long short-term memory.
\newblock {\em Neural Comput.}, 9:1735--1780, 1997.

\bibitem{Rumelhart}
D.~Rumelhart, G.~Hinton, and R.~Williams.
\newblock Learning representations by back-propagating errors.
\newblock {\em Nature}, 323:533–536, 1986.

\bibitem{NH2010}
V.~Nair and G.~E. Hinton.
\newblock Rectified linear units improve restricted boltzmann machines.
\newblock {\em In Proc. 27th International Conference on Machine Learning},
  pages 807--814, 2010.

\bibitem{kingma2014}
D.~P. Kingma and J.~Ba.
\newblock Adam: A method for stochastic optimization.
\newblock {\em {\rm arXiv:1412.6980}\hspace{-0.3em}}, 2014.

\bibitem{prechelt1998}
L.~Prechelt.
\newblock Automatic early stopping using cross validation: quantifying the
  criteria.
\newblock {\em Neural Netw.}, 11(4):761--767, 1998.

\bibitem{MFZNF2021}
M.~Morimoto, K.~Fukami, K.~Zhang, A.~G. Nair, and K.~Fukagata.
\newblock Convolutional neural networks for fluid flow analysis: toward
  effective metamodeling and low dimensionalization.
\newblock {\em Theor. Comput. Fluid Dyn.}, 35(5):633--658, 2021.

\bibitem{Fukamipump2022}
K.~Fukami, B.~An, M.~Nohmi, M.~Obuchi, and K.~Taira.
\newblock Machine-learning-based reconstruction of turbulent vortices from
  sparse pressure sensors in a pump sump.
\newblock {\em J. Fluids Eng.}, 144(12):121501, 2022.

\bibitem{LBBH1998}
Y.~LeCun, L.~Bottou, Y.~Bengio, and P.~Haffner.
\newblock Gradient-based learning applied to document recognition.
\newblock {\em Proc. IEEE}, 86(11):2278--2324, 1998.

\bibitem{pan2009survey}
S.~J. Pan and Q.~Yang.
\newblock A survey on transfer learning.
\newblock {\em IEEE Trans. Knowl. Data Eng.}, 22(10):1345--1359, 2009.

\bibitem{MFZF2020}
M.~Morimoto, K.~Fukami, K.~Zhang, and K.~Fukagata.
\newblock Generalization techniques of neural networks for fluid flow
  estimation.
\newblock {\em Neural Comput. Appl.}, 34:3647--3669, 2022.

\bibitem{guastoni2020convolutional}
L.~Guastoni, A.~G{\"u}emes, A.~Ianiro, S.~Discetti, P.~Schlatter, H.~Azizpour,
  and R.~Vinuesa.
\newblock Convolutional-network models to predict wall-bounded turbulence from
  wall quantities.
\newblock {\em J. Fluid Mech.}, 882:A27, 2021.

\end{thebibliography}

\end{document}